\documentclass[pdflatex,sn-mathphys-num]{sn-jnl}
\usepackage{amsmath}
\usepackage{flexisym}
\usepackage{listings}
\usepackage{amssymb,amsfonts}
\usepackage{caption}
\usepackage{supertabular,booktabs}
\usepackage{algorithmic}
\usepackage{graphicx}
\usepackage{booktabs}
\usepackage{subcaption}
\usepackage{textcomp}
\usepackage{placeins}
\usepackage{longtable}
\usepackage{xcolor}
\usepackage{ulem}
\usepackage{url}
\usepackage{textcomp}

\begin{document}


\title[Article Title]{\bf HiAER-Spike Software-Hardware Reconfigurable Platform for Event-Driven Neuromorphic Computing at Scale}

\author*[1]{\fnm{Gwenevere} \sur{Frank}}\email{jfrank@ucsd.edu}

\author[1,4]{\fnm{Gopabandhu} \sur{Hota}}\email{ghota@ucsd.edu}

\author[1]{\fnm{Keli} \sur{Wang}}\email{k3wang@ucsd.edu}

\author[1]{\fnm{Christopher} \sur{Deng}}\email{ckdeng@ucsd.edu}

\author[1]{\fnm{Krish} \sur{Arora}}\email{k7arora@ucsd.edu}

\author[1]{\fnm{Diana} \sur{Vins}}\email{dvins@ucsd.edu}

\author[1]{\fnm{Abhinav} \sur{Uppal}}\email{auppal@ucsd.edu}

\author[1]{\fnm{Omowuyi} \sur{Olajide}}\email{oolajide@ucsd.edu}

\author[1]{\fnm{Kenneth} \sur{Yoshimoto}}\email{kyoshimoto@ucsd.edu}

\author[1]{\fnm{Qingbo} \sur{Wang}}\email{qiw043@ucsd.edu}

\author[1,2]{\fnm{Mari} \sur{Yamaoka}}\email{myamaoka@ucsd.edu}

\author[1,3]{\fnm{Johannes} \sur{Leugering}}\email{jleugering@ucsd.edu}

\author[1]{\fnm{Stephen} \sur{Deiss}}\email{sdeiss@ucsd.edu}

\author*[1]{\fnm{Leif} \sur{Gibb$^\dagger$}}\email{lgibb@ucsd.edu}

\author*[1]{\fnm{Gert} \sur{Cauwenberghs$^\dagger$}}\email{gcauwenberghs@ucsd.edu}

\affil[1]{\orgdiv{Institute for Neural Computation}, \orgname{UC San Diego}, \orgaddress{\street{9500 Gilman Dr}, \city{La Jolla}, \postcode{92093}, \state{CA}, \country{USA}}}

\affil[2]{\orgname{Fujitsu}, \orgaddress{\street{4-1-1 Kamikodanaka}, \city{Nakahara-ku Kawasaki-shi}, \postcode{211-8588}, \state{Kanagawa}, \country{Japan}}}

\affil[3]{\orgname{Forschungszentrum Jülich}, \orgaddress{\street{52425}, \city{Jülich}, \country{Germany}}}

\affil[4]{\orgname{Qernel AI}, \orgaddress{\street{460 California Ave ste 204}, \city{Palo Alto}, \postcode{94306}, \state{CA}, \country{USA}}}

\abstract{In this work, we present HiAER-Spike, a modular, reconfigurable, event-driven neuromorphic computing platform designed to execute large spiking neural networks with up to 160 million neurons and 40 billion synapses - roughly twice the neurons of a mouse brain at faster than real time. This system, assembled at the UC San Diego Supercomputer Center, comprises a co-designed hard- and software stack that is optimized for run-time massively parallel processing and hierarchical address-event routing (HiAER) of spikes while promoting memory-efficient network storage and execution.
The architecture efficiently handles both sparse connectivity and sparse activity for robust and low-latency event-driven inference for both edge and cloud computing. A Python programming interface to HiAER-Spike, agnostic to hardware-level detail, shields the user from complexity in the configuration and execution of general spiking neural networks with minimal constraints in topology. The system is made easily available over a web portal for use by the wider community.
In the following, we provide an overview of the hard- and software stack, explain the underlying design principles, demonstrate some of the system's capabilities and solicit feedback from the broader neuromorphic community.  Examples are shown demonstrating HiAER-Spike's capabilities for event-driven vision on benchmark CIFAR-10, DVS event-based gesture, MNIST, and Pong tasks.
}



\maketitle

\noindent\textit{To appear in \textit{npj Unconventional Computing}}



\section{Introduction}
Spiking neural networks (SNNs) mirror the inherently event-driven way information is processed in the human brain by encoding it into the timing of \textit{spikes}. In the field of event-based sensing, particularly with dynamic vision sensors (DVS), this has proven to be a powerful \cite{deng2020rethinking} and highly energy efficient mode of computing. It seems likely that neuromorphic VLSI hardware could thus overcome some of the limitations of conventional von-Neumann computing architectures and provide similar power savings for other applications. But since there is still lacking hardware and software support for running large-scale SNN simulations, research has been limited to much smaller - and hence less capable - networks than those used in the field of Deep Learning.

\begin{figure}
\begin{center}
\textbf{System Overview}\par\medskip
\includegraphics[width=0.5\columnwidth]
  {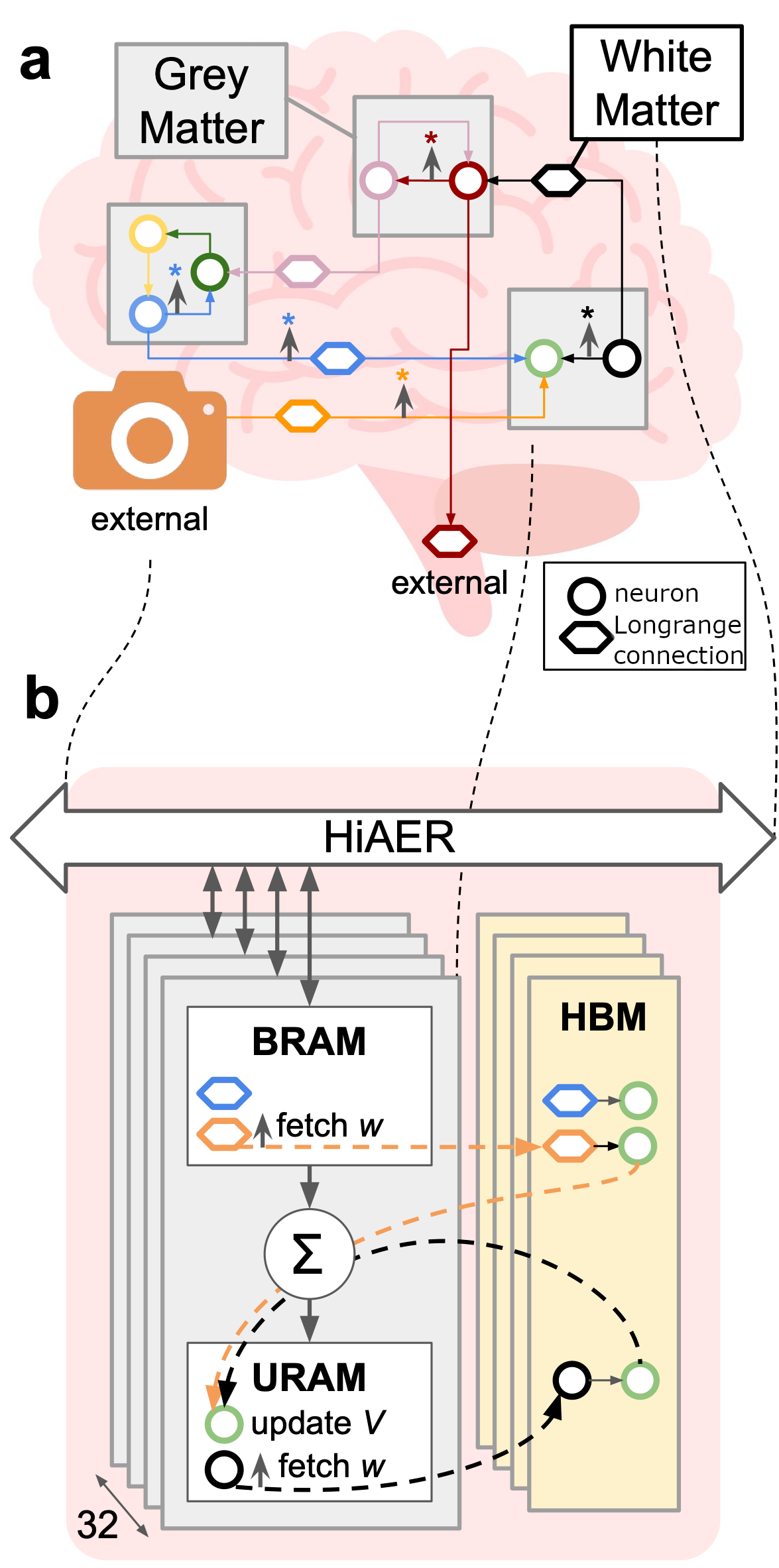}
  \caption{High-level system architecture of HiAER-Spike. (a) Neurons and synapses form the `grey matter' of dense local interconnects in the system while long-range connections (between cores and FPGAs) form the `white matter' of sparse global interconnects. (b) The hardware equivalent of (a) as implemented 
  in our multi-core architecture on the FPGA. The grey matter inside each core is implemented as sequentially updated integrate-and-fire neurons, whose internal state is stored in neural membrane registers in URAM, whereas spike events are routed through synaptic look-up tables in HBM. The white matter is implemented as a hierarchical multicast bus (HiAER) interconnecting axon spike register modules that are stored in BRAM across cores.
  }

  \label{fig:HiAER-Spike-top}
\end{center}
\end{figure}

To address this limitation, we have created a hierarchical address-event routing platform for spike-based processing (HiAER-Spike) for the explicit purpose of training and deploying SNNs at scale (Fig. \ref{fig:HiAER-Spike-top}). As many aspects of SNNs like neuron models and learning rules are still areas of active research, our system leverages a reconfigurable FPGA based computing architecture that allows for full customization and continuous improvements. Under the umbrella of the NSF-supported Computer and Information Science and Engineering (CISE) Community Research Infrastructure program, we make this system available to you, the research community, and the general public.

At present, we primarily target researchers focused on neuromorphic systems and SNNs while consolidating the system's features. In the next phase, we aim for wider adoption by a diverse cross section of users in the broader STEM research community, e.g. for large-scale brain simulations and the development of novel AI algorithms. To make the platform accessible to such a diverse audience, we created an intuitive and user-friendly open-source software interface that shields novice users from the challenges of operating and configuring such highly specialized neuromorphic hardware. Building on extensive existing network and storage infrastructure for user access and data sharing at the San Diego Supercomputer Center, the HiAER-Spike platform is hosted and maintained through the Neuroscience Gateway (NSG \cite{sivagnanam_neuroscience_2020}) Portal, which already serves over 1,100 registered users in the scientific community.

The insights and community feedback gathered from this system will inform the development of more targeted ultra-low power ASIC SNN accelerators in the final phase of the project.

In the remainder of this paper, we first give an overview of the hardware stack and how it accommodates large SNNs. Then we present the software interfaces that allow users to remotely access and utilize the system. Finally, using a single core
of one HiAER-Spike FPGA card, we demonstrate how the system operates on exemplary event-based vision use cases. We trained multilayer perceptrons (MLPs), LeNet-5, and spiking convolutional neural network (CNN) variants using PyTorch and converted the networks to run on the FPGA. The networks were implemented using binary and integrate-and-fire neurons. 

\section{Related Work}
To the best of our knowledge, no FPGA-based solution of the scale we are proposing has ever been made publicly available to the community before. Inspired by the human brain, there have been efforts by multiple research groups to develop dedicated hardware for accelerating SNNs.
Perhaps most similar in scope is SpiNNaker \cite{painkras2012spinnaker}. SpiNNaker's architecture is structured around ARM-based chips designed by the APT Advanced Processor Technologies Research Group from the University of Manchester. The largest system contains 32,400 chips and is capable of simulating hundreds of millions of neurons. The system is made available to researchers over a cloud interface. Although SpiNNaker is designed around custom chips that utilize a purpose-designed custom communication architecture, the actual computations are performed using ARM processor cores as opposed to custom-designed logic.
Intel's neuromorphic platform Loihi 2 is designed around a custom architecture \cite{davies2021taking}. Loihi 2 chips consist of six processing units and 128 neuron cores per chip. Neuron cores allow for the specification of custom neuron models using a provided microcode. Loihi 2 supports up to a maximum of 1 million synapses per chip. Access to Loihi 2 devices is provided over a cloud interface to members of Intel's neuromorphic research community. BrainScaleS is a mixed-signal neuromorphic chip simulating 180,000 neurons and 40 million synapses. The upgraded platform, BrainScaleS-2, supports more complex neuron and synapse models \cite{pehle2022brainscales}.
IBM recently introduced their NorthPole chip, an evolution of their True North design \cite{modha2023neural}. NorthPole is a custom architecture developed by IBM and supports up to 1,048,576 neurons per chip. NorthPole does not appear to be currently accessible to the public.
In comparison to existing work, HiAER-Spike is designed to leverage the advantage of having a widely available system based on reconfigurable hardware rather than fixed ASIC designs or more general microprocessor cores, in order to rapidly incorporate user feedback on requested features into new revisions of the digital hardware designs in the form of new FPGA bitstreams and provide a community-driven testbed for iterating on new digital neuromorphic hardware designs.

\section{Hardware System Organization and Software Co-design}

The architecture of the HiAER-Spike system hosted at the San Diego Supercomputer Center (SDSC) consists of 6 servers, each with two 32 core AMD EPYC processors, 1 TB of shared DRAM, and over 29 TB of enterprise-grade SSDs. One server is the head node for the system and interfaces with the Neuroscience Gateway (NSG) to allow users from anywhere in the world to run jobs on HiAER-Spike hardware at SDSC through the NSF ACCESS supercomputing network (NSF Program 21-555). The other 5 servers each contain 8 Alpha Data ADM-PCIE-9H7 FPGA boards that provide terabits of I/O bandwidth, along with 460 GBPS memory bandwidth to 8 GB of high bandwith memory (HBM) on each FPGA chip plus extensive on-chip SRAM and programmable logic using the Xilinx XCVU 37P chip. The memory bandwidth is spread across 32 parallel processing cores in each FPGA that each can emulate multiple neuron models. There are 3 Arista switches  that allow server-to-server spike messaging, server management, file sharing, and job management. Each FPGA board also has 4 bidirectional 1 Tbps FireFly connections for passing spikes between boards within each server. Server CPUs coordinate with the FPGA processing cores via PCIe 3.0 to program network definitions and execute synaptic weight updates for various learning algorithms. 

Each FPGA, in turn, contains multiple SNN cores that will run fully in parallel with dedicated interfaces to the on-module HBM. Multiple layers of multicast address-event routing schemes (Network on Chip, FireFly, and Ethernet) \cite{park2016hierarchical, iscas_hiearahb} will allow spikes to travel efficiently between cores, FPGAs, and servers.
Across all these levels of the hierarchy, the system will track spike events with 1 ms resolution and support synaptic learning algorithms that require careful accounting for time differences between pre- and postsynaptic spikes, such as variations of spike-timing-dependent plasticity (STDP).

This paper presents initial benchmarks using only a single operational core on one FPGA, and outlines how we intend to expand this up to 1280-fold to multiple cores per FPGA across 40 FPGAs. We aim for each FPGA to support up to four million neurons and one billion synapses, yielding a total capacity of up to 160 million neurons and 40 billion synapses for the entire system - more than twice the number of neurons of a typical mouse brain \cite{herculano2006cellular} and at faster-than real-time simulation speed.

To be able to deploy networks at such scale, we have developed a network partitioning and resource allocation algorithm that assigns SNN simulation jobs to servers, FPGA boards, and cores as required to meet the user’s requirements \cite{mysore_hierarchical_2022}.

\section{Storing Networks}

\begin{figure}
\begin{center}
\textbf{Core Architecture}\par\medskip
  \includegraphics[scale=0.16] 
  {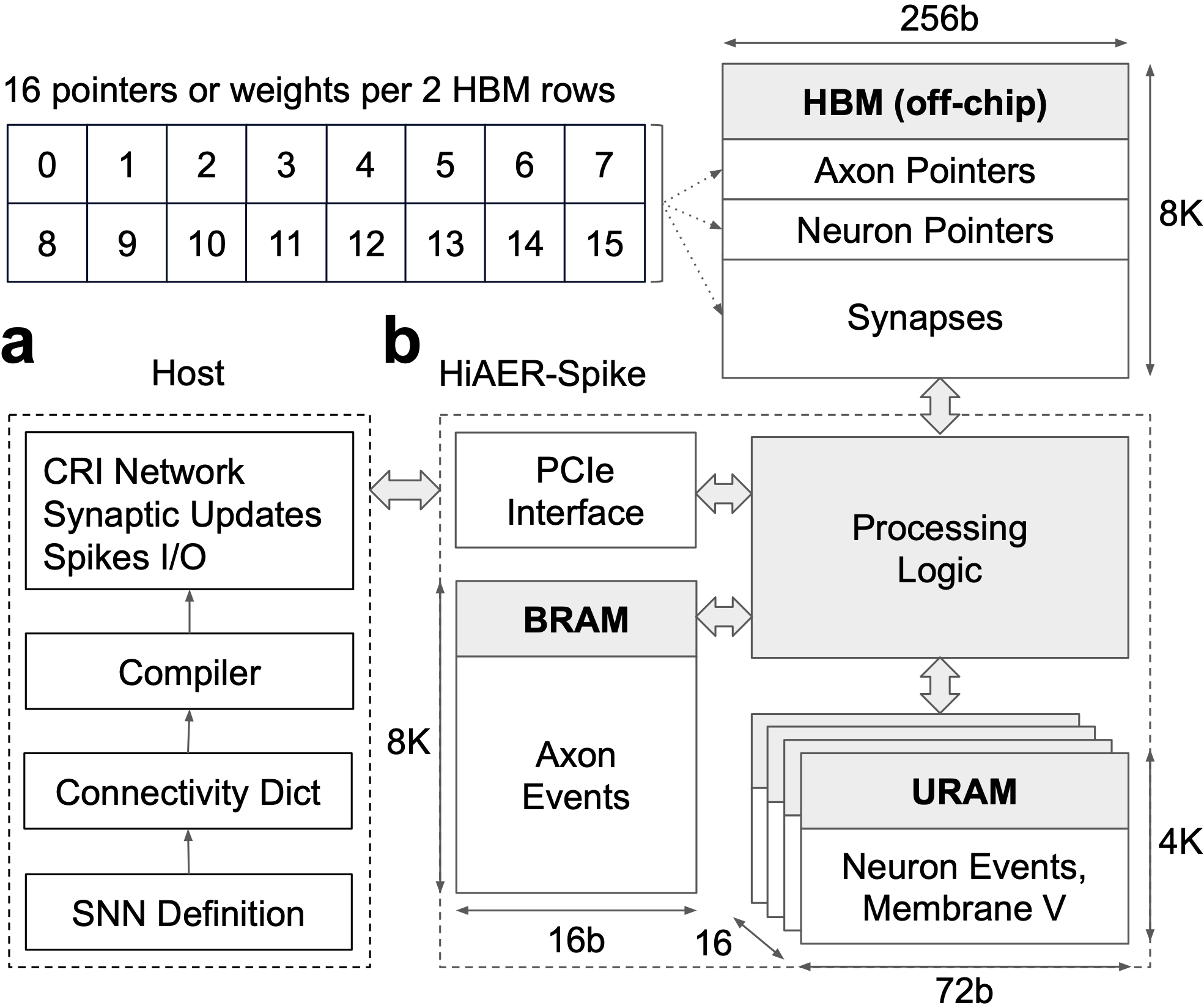}
  \caption{(a) Host programming interface with SNN compiler and low-level hardware interface. (b) Heterogeneous memory organization within a single core, as well as the off-chip HBM storing the synaptic connectivity table. On-chip URAM and BRAM store state variables of axons and neurons. The top-left panel shows the layout of the data structure in HBM, supporting parallelism of 16 neurons per single core.}

  \label{fig:process-flow}
\end{center}
\end{figure}

For large-scale neuromorphic hardware, synaptic storage density is an important determiner of efficiency and the maximum supported network connectivity. Currently deployed neural networks are frequently sparse and can be made sparser by pruning and quantization techniques. For this reason, we opt to use an adjacency list to store network connectivity, rather than a cross-bar structure\cite{kulkarni2019neuromorphic,correll2020fully}, as adjacency lists store sparse networks highly efficiently \cite{pedroni2019memory}.

The network is stored inside the HBM in a format that includes pointers for neurons and axons (inputs to the network) that point to the respective synaptic weights. A portion of memory is reserved for axon and neuron pointers, and another portion is reserved for synapse definitions. Each pointer consists of a starting address and a number of rows in HBM that defines the region in memory where the outgoing synapses from the pointer's corresponding neuron or axon are defined.
The HBM, with 8GB capacity per FPGA card, is divided into segments of 16 slots spanning two rows (Fig. \ref{fig:process-flow}) with each slot storing a single pointer or synapse value. The network compiler is made aware of the memory alignment constraints of HBM, that is that synapses must utilize the same slot number as the pointer corresponding with their postsynaptic neuron, and adjusts the neuron and axon assignments to obtain maximum packing density in HBM, lowering execution latency. Having the presynaptic neuron pointer store just the base (start) address and the number of rows of HBM occupied by the postsynaptic connections, as opposed to absolute addresses, further reduces memory usage. Neuron pointers are grouped by their corresponding neuron model in memory.

The routing of spikes then proceeds in two phases: first, for each neuron that fired in the current time step and for each incoming externally driven axon, the pointers to all postsynaptic connections are read into a queue. In the second phase, postsynaptic neuron addresses and the number of rows each neuron/axon's synapses span are retrieved for the enqueued pointers, corresponding synaptic weights are fetched from HBM, and the membrane potentials $V_i$ of the postsynaptic neurons are updated - possibly generating new spikes for the next cycle - and so on.

HBM allows accessing large packets over multiple ports, which we leverage to parallelize event lookups in the first phase and membrane potential updates in the second phase.
We use local on-chip Block RAM (BRAM) and UltraRAM (URAM), laid out in the same structure as the HBM, to store spike events and membrane potentials, because they are queried every time step and thus have a dense access pattern. This hybrid approach, combining HBM and on-chip SRAM, provides significant energy and latency improvements. Further details of the hardware microarchitecture will be discussed in a future manuscript.

\section{Software Description}

We provide a software interface to allow end users to easily utilize the hardware. The interface is implemented in C++ and Python and integrates with a community-developed Python library, named hs\_api, for defining and running networks on the hardware, which emerged out of a workgroup topic at the 2022 Telluride Neuromorphic Cognition Engineering Workshop (\url{https://sites.google.com/view/telluride-2022/home}).

\subsection{Supported neuron models}
\label{nmodels}
HiAER-Spike supports networks composed of multiple types of neurons, as displayed in Table \ref{tab:neuronType}. Currently, simple binary neurons (referred to as ANN neurons), that is, neurons that either spike or do not spike at each time step and accumulate no membrane potential between steps, and leaky-integrate-and-fire (LIF) neurons, are supported with an optional addition of noise to the membrane potential at each time step for both models. Networks may be composed of multiple types of stochastic and deterministic LIF and binary neurons with different parameters. ANN neurons have two parameters (threshold $\theta$ and noise shift $\nu$) and each LIF neuron has three parameters  ($\theta$, $\nu$, and leak parameter $\lambda$). Threshold $\theta$ determines the membrane potential at which the neuron spikes and the membrane potential is reset to zero. Noise shift $\nu$ is a 6-bit signed integer that controls the magnitude of random noise added to the membrane potential at each step. Noise is a 17-bit signed integer randomly generated and then right shifted by $\nu$ if $\nu$ is negative or left shifted by $\nu$ if $\nu$ is positive. The least significant bit of the noise is always set to one to balance the distribution around zero. Finally, $\lambda$ controls the voltage leakage that occurs during each time step $voltage=voltage-voltage/2^{\lambda}$. Each neuron in a network can be assigned a corresponding neuron model with no restrictions.

\begin{table}[h]
\caption*{\textbf{Neuron Models}}
\centering
\begin{tabular}{lllll}
\hline
\multicolumn{1}{|l|}{\begin{tabular}[c]{@{}l@{}}Neuron\\ Model\end{tabular}} & \multicolumn{1}{l|}{Parameters} & \multicolumn{1}{l|}{Noise Update} & \multicolumn{1}{l|}{Spike Update} & \multicolumn{1}{l|}{Membrane Update} \\ \hline
LIF & \begin{tabular}[c]{@{}l@{}}$\theta, \nu,\lambda$\end{tabular} & \begin{tabular}[c]{@{}l@{}}$V_i = V_i + \xi$, \\ where $\xi \sim U_{(-2^{16}, 2^{16})}\cdot2^{\nu}$ \end{tabular} & \begin{tabular}[c]{@{}l@{}}$S_i = (V_i > \theta)$\\ $S_i \to V_i=0$\end{tabular} & $V_i=V_i-V_i/2^{\lambda} + \sum_j w_{ij}S_j $ \\[10pt]
\begin{tabular}[c]{@{}l@{}}ANN\\(binary)\end{tabular} & $\theta, \nu$ & \begin{tabular}[c]{@{}l@{}}$V_i = V_i + \xi$, \\ where $\xi \sim U_{(-2^{16}, 2^{16})}\cdot2^{\nu}$ \end{tabular} & \begin{tabular}[c]{@{}l@{}}$S_i = (V_i > \theta)$\\ $S_i \to V_i=0$\end{tabular} & $V_i= \sum_j w_{ij}S_j$
\end{tabular}
\caption{Parameters for the two supported classes of neuron model in the software API are shown. First a noise update is applied to the membrane potential, then the neuron's membrane potential is thresholded for spikes and reset to zero if a spike is detected, and finally a membrane update is applied and incoming synaptic events are integrated. $U_{(-2^{16}, 2^{16})}\cdot2^{\nu}$ in this table represents the \textit{modified} discrete uniform distribution of odd integers implied by the noise generation method of Section \ref{nmodels}. The LIF neuron can be reduced to an approximation of an integrate-and-fire (IF) neuron by setting $\lambda$ to its maximum supported value of $2^{6}-1$. If $\nu$ is set to a value greater than -17 for an ANN (binary) neuron, it will function as a Boltzmann-like binary stochastic neuron.}
\label{tab:neuronType}
\end{table}

\subsection{API}
The software exposes an API that allows users to define, run and interact with SNNs on the hardware. Networks can be defined by using a collection of simple Python objects. Complex architectures can be built programatically from these simple data structures, or alternatively, conversion code exists to implement more complex machine learning architectures developed with tools like PyTorch. This section will briefly describe the software primitives provided by the API. 

The \lstinline[language=Python]{LIF_neuron} and \lstinline[language=Python]{ANN_neuron} classes can be used to define a neuron model. The user must supply an \lstinline[language=Python]{axons} dictionary. Axons represent user-controllable input from the external world into the network, through which a user may send a spike to interact with multiple postsynaptic neurons. The user must also supply a \lstinline[language=Python]{neurons} dictionary. Finally, the user must supply an outputs list. This is a list of all the unique keys of neurons whose spiking activity the user wishes to monitor.

Once the necessary axons dictionary, neurons dictionary, and outputs list are defined, they may be passed to the constructor to create an instance of the 
\lstinline[language=Python]{CRI_network} class that exposes functions through which the user can interact with the network. The \lstinline[language=Python]{step} method can be used to run one timestep of the network. The methods \lstinline[language=Python]{read_synapse} and \lstinline[language=Python]{write_synapse} can be used to change synapse weights after calling the 
\lstinline[language=Python]{CRI_network} constructor.

\subsection{Running Inference}
The hs\_api library enables users to define and interact with networks either in software simulation on their local machine, or accelerated on the HiAER-Spike hardware through NSG. The API of the hs\_api library remains exactly the same in both cases. If the user is running on their local machine where the HiAER-Spike hardware is not detected, network inference is run using a simulation of the hardware operations implemented in Python. If the user is running their code on the HiAER-Spike cluster over NSG, the hs\_api library detects the presence of the HiAER-Spike hardware and runs the inference on the accelerator hardware. This makes a seamless transition possible for users who want to develop on their local devices and then submit larger workloads to run on the HiAER-Spike cluster.

The \textit{simulator} currently implements inference using sparse matrix operations and fixed-bit integer arithmetic. The network is represented by two sparse matrices holding the weights for axons and neurons, respectively. During the simulation, the membrane potentials are updated according to the noise, leakage, and membrane reset behavior specified by the chosen neuron model of each neuron. Then two binary vectors are constructed, holding the indices of all axons or neurons, respectively, that yielded a spike in that timestep. These vectors are multiplied by the axon and neuron weight matrices, respectively, to calculate the total input for each neuron in the network, and the membrane potentials are updated accordingly. Finally, an output spike is recorded at the current timestep for any output neuron that just fired. An excerpt of the core components of the simulator is shown in the Supplementary Information, although some details, such as provisions for fixed-bit arithmetic, are omitted for readability.

\section{Examples for Event-Based Vision}\label{results}

\begin{table*}[]
\begin{center}
\caption*{\textbf{Accuracy, Latency, and Energy of HiAER-Spike}}
\fontsize{7pt}{9pt}\selectfont
\setlength{\tabcolsep}{2pt}
\hspace*{-0.5in}
\begin{tabular}{lllllllllllllll}
\hline
\multicolumn{1}{|l|}{\begin{tabular}[c]{@{}l@{}}Model \\ Type\end{tabular}} & \multicolumn{1}{l|}{\begin{tabular}[c]{@{}l@{}}Input\\ Shape\end{tabular}} & \multicolumn{1}{l|}{Architecture} & \multicolumn{1}{l|}{Task} & \multicolumn{1}{l|}{Axons} & \multicolumn{1}{l|}{Neurons} & \multicolumn{1}{l|}{Weights} & \multicolumn{1}{l|}{\begin{tabular}[c]{@{}l@{}}Software\\ Acc. (\%)\\or Score\end{tabular}} & \multicolumn{1}{l|}{\begin{tabular}[c]{@{}l@{}}HiAER\\ Acc. (\%)\\or Score\end{tabular}} & \multicolumn{1}{l|}{\begin{tabular}[c]{@{}l@{}}HBM\\ Energy\\ ($\mu$J) \end{tabular}} & \multicolumn{1}{l|} {\begin{tabular}[c]{@{}l@{}}Latency\\ ($\mu$s) \end{tabular}}\\ \hline
MLP & (1, 28, 28) & 128→10 & \begin{tabular}[c]{@{}l@{}}MNIST\end{tabular} & 784 & 138 & 101,632 & 96.59 & 96.59 & 1.1$\pm$0.3 & 4.2$\pm$0.6 \\[1pt]
MLP & (1, 28, 28) & \begin{tabular}[c]{@{}l@{}}2k→1k→10\end{tabular} & \begin{tabular}[c]{@{}l@{}}MNIST\end{tabular} & 784 & 3,010 & 3,578,000 & 97.66 & 97.66 & 19.3$\pm$3.8 & 45.5$\pm$8.3 \\[2pt]
\begin{tabular}[c]{@{}l@{}}LeNet-5\\ (stride=2)\end{tabular} & (1, 28, 28) & \begin{tabular}[c]{@{}l@{}}C(6)→C(16)\\→3FC\end{tabular} & \begin{tabular}[c]{@{}l@{}}MNIST\end{tabular} & 784 & 1,334 & 44,190 & 97.76 & 97.76 & 6.4$\pm$1.1 & 18.9$\pm$2.5 \\[7pt]
\begin{tabular}[c]{@{}l@{}}LeNet-5\\ (max pool)\end{tabular} & (1, 28, 28) & \begin{tabular}[c]{@{}l@{}}C(6)→MP\\ →C(16)→MP\\→3FC\end{tabular} & \begin{tabular}[c]{@{}l@{}}MNIST\end{tabular} & 784 & 5,814 & 44,190 & 98.14 & 98.14 & 17.1$\pm$2.8 & 48.6$\pm$5.9 \\[12pt]
\begin{tabular}[c]{@{}l@{}}Spiking\\ CNN\end{tabular} & (2, 63, 63) & \begin{tabular}[c]{@{}l@{}}C(1)→3FC\end{tabular} &  \begin{tabular}[c]{@{}l@{}}DVS\\ Gesture\end{tabular} & 7,938 & 1,115 & 119,054 & 55.47 & 54.51 & 79.8$\pm$22.5 & 184.9$\pm$48.4 \\[7pt]
\begin{tabular}[c]{@{}l@{}}Spiking\\ CNN\end{tabular} & (2, 63, 63) & \begin{tabular}[c]{@{}l@{}}3C(100)→3FC\end{tabular} & \begin{tabular}[c]{@{}l@{}}DVS\\ Gesture\end{tabular} & 7,938 & 109,615 & 816,004 & 64.45 & 64.58 & 3,268.1$\pm$756.0 & 7,326.4$\pm$1,623.2\\[7pt]
\begin{tabular}[c]{@{}l@{}}Spiking\\ CNN\end{tabular} & (2, 90, 90) & \begin{tabular}[c]{@{}l@{}}C(6)→C(16)\\→3FC\end{tabular} & \begin{tabular}[c]{@{}l@{}}DVS\\ Gesture\end{tabular} & 16,200 & 17,709 & 781,704 & 68.75 & 68.75 & 510.7$\pm$145.4 & 1,156.2$\pm$312.5 \\[7pt]
\begin{tabular}[c]{@{}l@{}}Spiking\\ CNN\end{tabular} & (15, 32, 32) & \begin{tabular}[c]{@{}l@{}}C(16)→2C(100)\\ →2FC\end{tabular} & \begin{tabular}[c]{@{}l@{}}CIFAR-10\end{tabular} & 15,360 & 38,122 & 1,954,880 & 82.56 & 80.49 & 4,770.7$\pm$614.3 & 10,508.5$\pm$1,319.2\\[7pt]
\begin{tabular}[c]{@{}l@{}}Spiking\\ CNN\end{tabular} & (2, 84, 84) & \begin{tabular}[c]{@{}l@{}}C(32)→C(64)\\ →C(64)→2FC\end{tabular} & \begin{tabular}[c]{@{}l@{}}DVS\\ Pong\end{tabular} & 14,112 & 21,638 & 1,682,432 & 20.74 & 20.36 & 149.3$\pm$30.3 & 425.7$\pm$64.9 
\end{tabular}
\end{center}
\caption{Accuracy, latency, and energy performance of HiAER-Spike on various event-based vision network benchmarks. Software accuracy is given for models after weight quantization.
\textit{Score}: mean score in 50 episodes of DVS-converted Atari Pong (maximum possible is 21). \textit{HBM Energy}: energy per inference (mean$\pm$SD) calculated from HBM accesses reported by the FPGA. \textit{Latency}: latency per inference (mean$\pm$SD) calculated from clock cycles reported by the FPGA.}
\label{tab:model_results}
\end{table*}

\begin{figure}
\begin{center}
\textbf{DVS Waving Gesture}\par\medskip
  \includegraphics[scale=0.3]{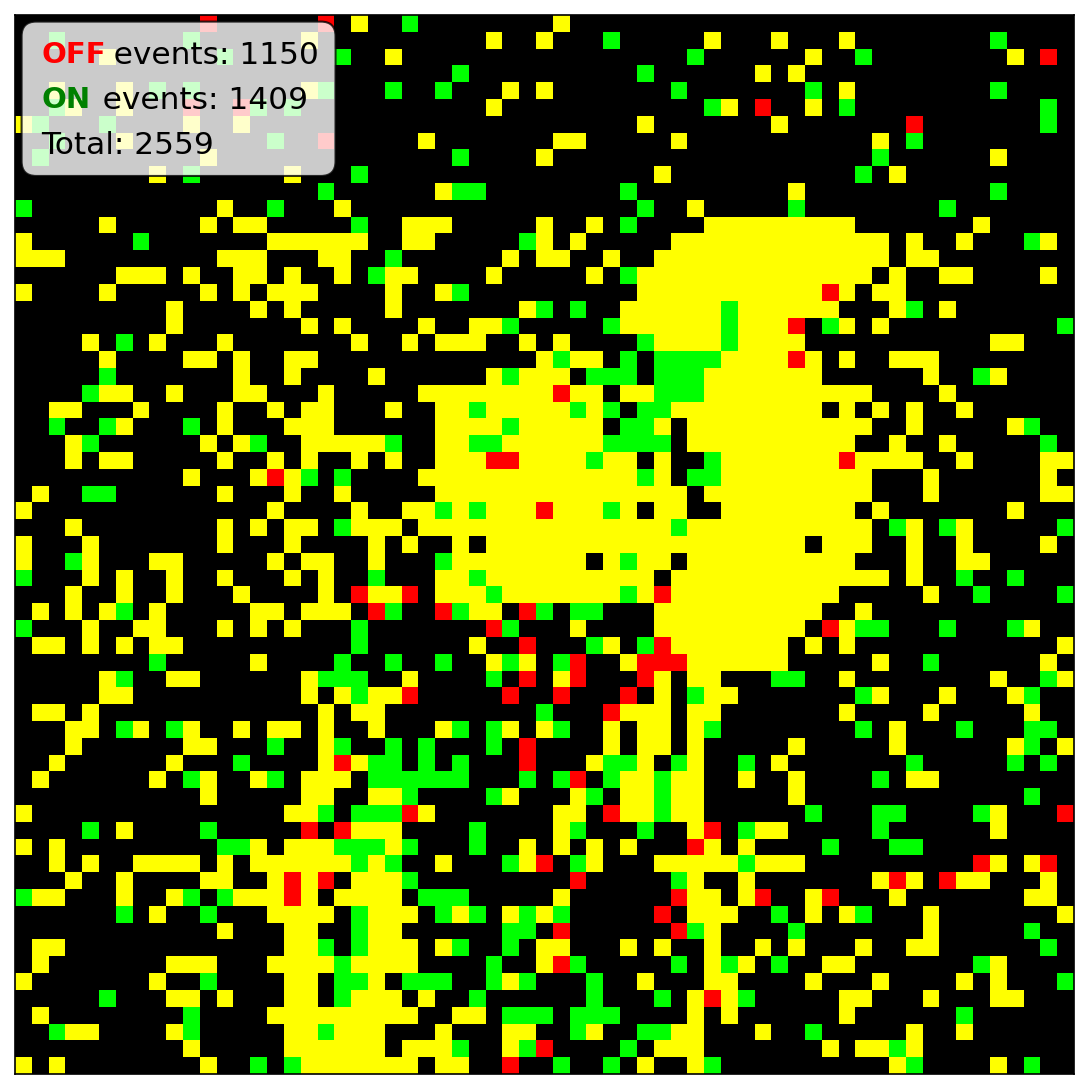}
  \caption{DVS Gesture example.  Representation of a waving gesture frame processed into two 63×63 pixel channels corresponding to ON and OFF events, with overlapped visualization showing ON events (green), OFF events (red), and overlapping ON and OFF events (yellow). One of ten frames accumulated across a gesture.}
  \label{fig:dvswave}
\end{center}
\end{figure}

\begin{figure}
\begin{center}
\textbf{DVS Pong}\par\medskip
  \includegraphics[scale=.3]{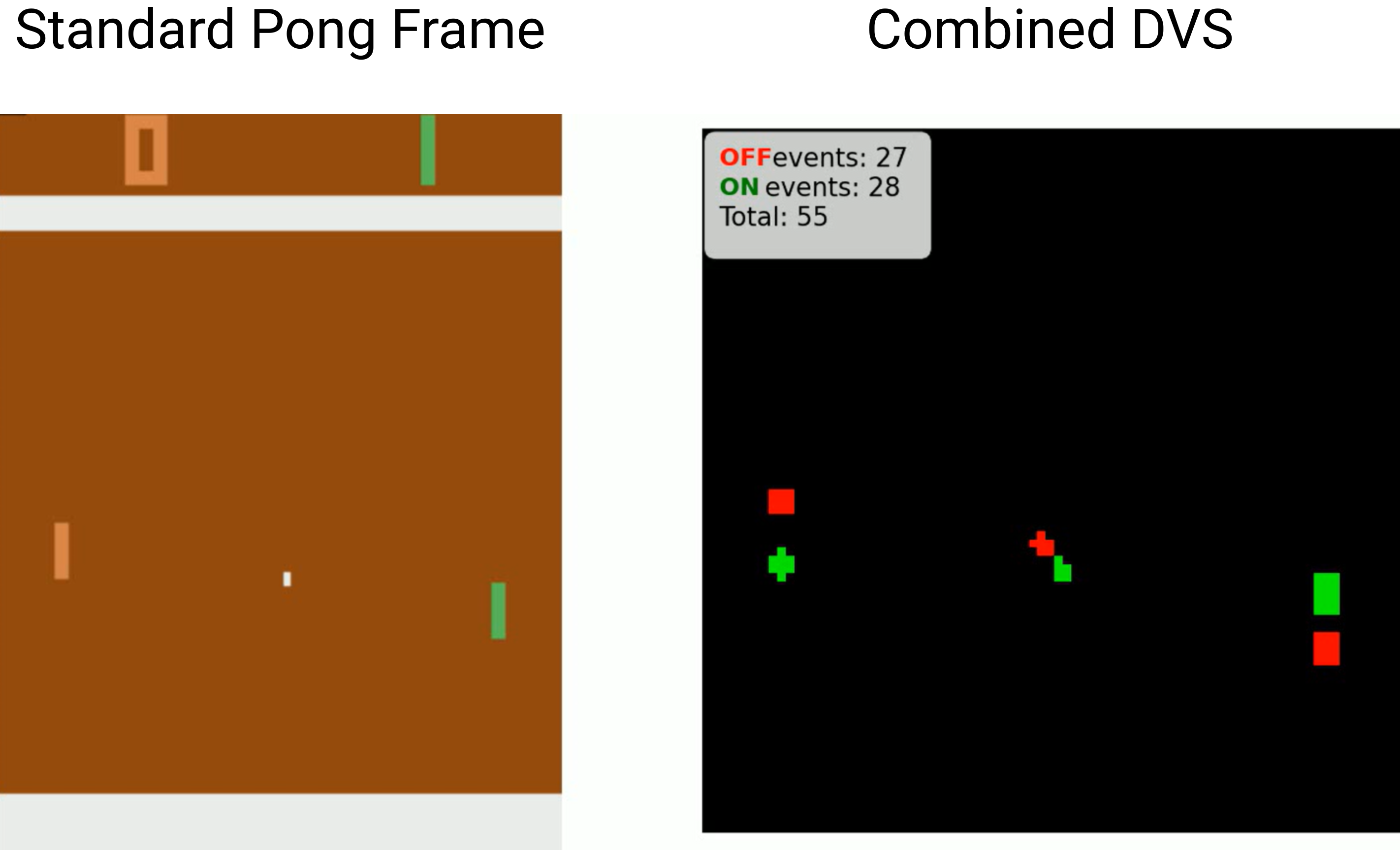}
  \caption{A 160x210-pixel Atari Pong frame (left) and the DVS representation of that frame (right) processed into two 84×84-pixel channels corresponding to ON and OFF events, with overlapped visualization showing ON events (green) and OFF events (red), detecting 27 OFF and 28 ON events in the current frame. The location of green ON pixels relative to red OFF pixels indicates the direction of motion (e.g. the ball in the center is moving diagonally towards the lower right).}
  \label{fig:pong}
  \end{center}
\end{figure}

 \begin{figure*}
 \begin{center}
 \textbf{DVS Gesture SNN Size and Quantization}\par\medskip
   \includegraphics[scale=.45]{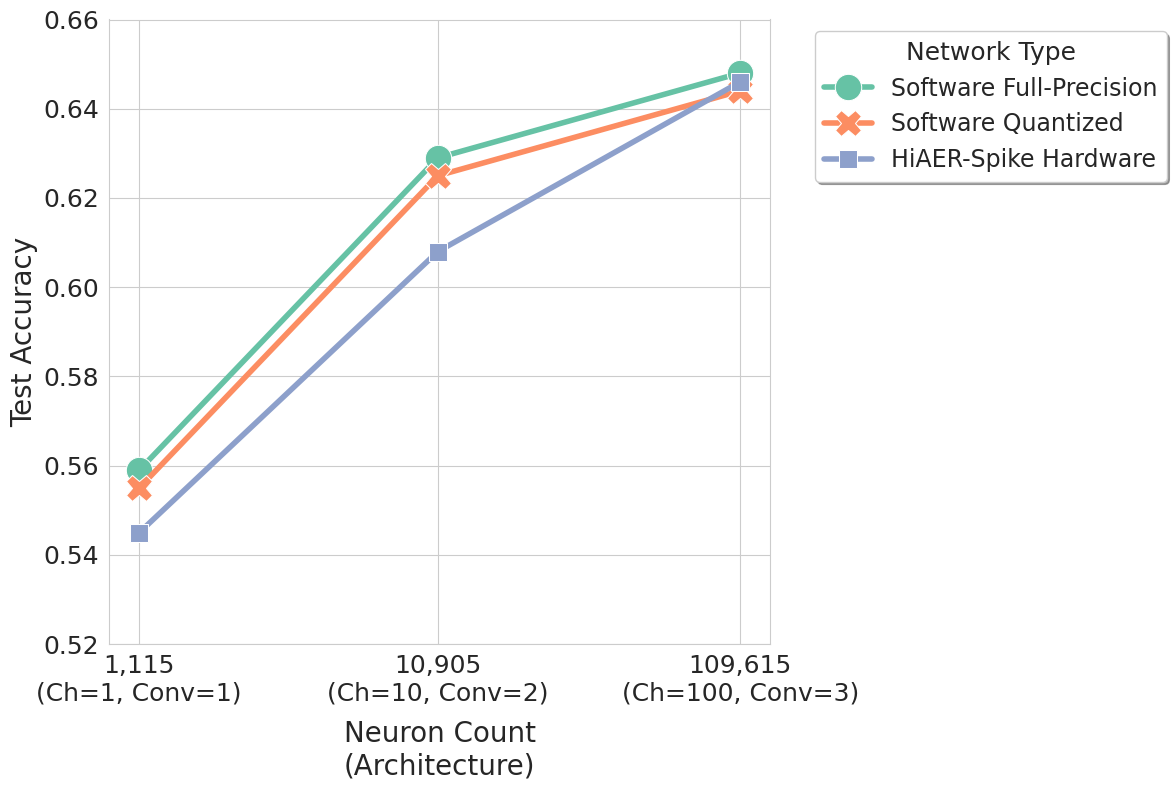}
   \caption{Test accuracy of the DVS gesture network compared across different model sizes and full-precision software models, quantized software models, and hardware.}
   \label{fig:dvs_gesture_result}
   \end{center}
 \end{figure*}

\begin{table}[]
\centering
\caption*{\textbf{MNIST Comparison Across Neuromorphic Platforms}}

\begin{tabular}{lllll}
\hline
\multicolumn{1}{|l|}{System} & \multicolumn{1}{l|}{\begin{tabular}[c]{@{}l@{}}Model Size\\ (Neurons)\end{tabular}} & \multicolumn{1}{l|}{\begin{tabular}[c]{@{}l@{}}Accuracy\\ (\%)\end{tabular}} & \multicolumn{1}{l|}{\begin{tabular}[c]{@{}l@{}}Energy \\ ($\mu J$)\end{tabular}} & \multicolumn{1}{l|}{\begin{tabular}[c]{@{}l@{}}Latency\\ ($\mu s$)\end{tabular}} \\ \hline
HiAER-Spike & 138 & 96.59 & 1.1 & 4.2 \\
HiAER-Spike & 5,814 & 98.14 & 17.1 & 48.6 \\
Loihi \cite{lenz2023ultra} & 5,400 & 99.23 & 182.46 & 4,900 \\
SpiNNaker \cite{stromatias2015scalable} & 1,790 & 95.01 & N/A & 20,000 \\
TrueNorth \cite{esser2015backpropagation} & 7,680{*} & 99.42 & 108 & N/A
\end{tabular}
{\footnotesize
{*}Calculated from number of neuron cores utilized and neurons per core.\\
\textit{First row}: HiAER-Spike model of lowest HBM energy and latency.\\
\textit{Second row}: HiAER-Spike model of highest accuracy.\\
\textit{Energy} and \textit{Latency}: per inference.\\
HiAER-Spike energy and latency are mean per inference calculated from HBM accesses and clock cycles, respectively, reported by the FPGA.\\
HiAER-Spike MNIST models use ANN (binary) neurons.\\ 
\textit{N/A}: data not available.
}
\caption{Performance metrics for HiAER-Spike and multiple other neuromorphic platforms running MNIST digit classification models.
}
\label{tab:mnistcomp}

\end{table}

\begin{table}[]
\caption*{\textbf{DVS Gesture Comparison Across Neuromorphic Platforms}}
\centering
\begin{tabular}{lllll}
\hline
\multicolumn{1}{|l|}{System} & \multicolumn{1}{l|}{\begin{tabular}[c]{@{}l@{}}Model Size\\ (Neurons)\end{tabular}} & \multicolumn{1}{l|}{\begin{tabular}[c]{@{}l@{}}Accuracy\\ (\%)\end{tabular}} & \multicolumn{1}{l|}{\begin{tabular}[c]{@{}l@{}}Energy\\ ($\mu J$)\end{tabular}} & \multicolumn{1}{l|}{\begin{tabular}[c]{@{}l@{}}Latency\\ ($\mu s$)\end{tabular}} \\ \hline
HiAER-Spike & 1,115 & 54.51 & 79.8 & 184.9 \\
HiAER-Spike & 17,709 & 68.75 & 510.7 & 1,156.2 \\
Loihi \cite{massa2020efficient} & N/A & 89.64 & N/A & 11,430 \\
SpiNNaker2 \cite{arfa2025efficient} & 9,907 & 94.13 & 459,000 & N/A  \\
TrueNorth \cite{amir2017low} & N/A & 96.49 & 18,700 & 104,600
\end{tabular}
\footnotesize{
\textit{First row}: HiAER-Spike model of lowest HBM energy and latency.\\
\textit{Second row}: HiAER-Spike model of highest accuracy.\\
\textit{Energy} and \textit{Latency}: per inference.\\
HiAER-Spike energy and latency are mean per inference calculated from HBM accesses and clock cycles, respectively, reported by the FPGA.\\
HiAER-Spike DVS Gesture models use LIF neurons with membrane time constant $2^{63}$.\\ 
\textit{N/A}: data not available.
}
\caption{Performance metrics for HiAER-Spike and multiple other neuromorphic platforms running DVS Gesture classification spiking neural networks.}
\label{tab:dvscomp}
\end{table}

Here, we present results for a selection of exemplary use cases, using only a single core of one FPGA card. We implemented a series of PyTorch and SpikingJelly models to test and demonstrate our HiAER-Spike FPGA hardware design, software interface, and conversion methods. The complexity of the models and tasks ranged from foundational MLPs and CNNs (variants of the LeNet-5 architecture) trained on binarized MNIST, to more complex spiking CNNs trained on the DVS Gesture or CIFAR-10 datasets or on an implementation of the classic video game \textit{Pong} utilizing a DVS-based input representation.

First, we implemented MLP and LeNet-5 variants with ANN (binary) neurons. The networks were trained on binarized MNIST, a dataset consisting of 28x28-pixel binary images of digits from 0 to 9, using quantization-aware training with binarized sigmoidal activation functions and int16 weights. These networks were trained in PyTorch using a learning rate of 0.001, a batch size of 64, and usually an early-stopping patience of 5 epochs. The LeNet-5 variant with max pooling used an early-stopping patience of 20. The model weights that achieved the lowest validation loss were saved.

We evaluated the software accuracy of the networks by running inference on the test dataset in PyTorch. We then converted each network to run on HiAER-Spike as described in Supplementary Section A.2. Each image in the test set was input into the network as a set of axon activations during one time step. The signal from one image was allowed to propagate to the output layer before the next image was passed into the network. There are 10 neurons in the output layer, one neuron for each of the digits 0 to 9. The output neuron with the highest membrane potential after each image presentation is the model’s prediction.

Table \ref{tab:model_results} shows the software and HiAER-Spike accuracies for a selection of MLP and LeNet-5 variants. All LeNet-5 variants and the larger MLP variant achieved over 97\% test accuracy. All MLP and LeNet-5 variants with input size (1, 28, 28) achieved identical test accuracies on PyTorch and HiAER-Spike, validating the conversion pipeline and capabilities of the hardware. The only one of the six MLP and LeNet-5 variants that we trained whose HiAER-Spike test accuracy fell short of perfectly matching its PyTorch test accuracy was one whose MNIST inputs were resized to (1, 90, 90), and its discrepancy was only 0.01\% (not included in Table \ref{tab:model_results}).

The hardware's energy usage is primarily dominated by HBM accesses; thus energy consumption was approximated by the product of the energy cost of a single HBM access and the number of HBM accesses performed during an inference.

Second, in SpikingJelly, we trained a set of spiking CNNs consisting of different numbers of convolutional layers (5 x 5 kernels, stride 2) with different numbers of output channels, followed by flattening and 3 linear layers (120 units, 84 units, 11 output units). For our spiking layers, we used a modified SpikingJelly LIFNode that matches HiAER-Spike's spike threshold method ($>$ rather than $\ge$) and order of operations within a timestep (the axonal and neuronal inputs are added to the membrane potential at the end of the timestep rather than the beginning). In both SpikingJelly and HiAER-Spike, we set the membrane time constant to $2^{63}$ to approximate an integrate-and-fire neuron without a leakage ($\lambda$) term, and we used a hard reset to 0 and no synaptic current decay.

We trained these networks on the IBM DVSGesture dataset \cite{amir2017low}, which contains clips of 11 different gestures (corresponding to our network’s 11 outputs) under 3 different lighting conditions recorded using a DVS camera. Fig. \ref{fig:dvswave} shows an example of the event data for a gesture accumulated into a frame. We split each gesture into 10 frames corresponding to 10 timesteps, following the methodology of SpikingJelly’s official DVSGestureNet tutorial (\url{https://spikingjelly.readthedocs.io/zh-cn/latest/activation_based_en/14_classify_dvsg.html}).

We preprocessed the DVS events into binarized frames of size (63, 63) or (90, 90) for input to the spiking CNN, and used two channels to separately represent ON and OFF events. The networks were trained using an ATan surrogate gradient function and a cosine annealing learning rate scheduler starting with a learning rate of 0.001 and reaching a minimum learning rate of 0 at 100 epochs. We used a batch size of 64 and an early-stopping patience of 20 epochs. After each timestep, the spike counts for the 11 output neurons were recorded. The total spike counts were then divided by the number of timesteps to obtain a spike rate for each neuron. The highest spike rate was used to pick a gesture.

The weights were then quantized to 16-bit integers, and the resulting networks were converted into the format needed for HiAER-Spike (using methods based closely on those described in Supplementary Section A.2) and deployed on the hardware. An axon map was constructed to map the 2-channel input data to the first convolutional layer. We repeated the process for each layer, and HiAER-Spike output neurons were defined for the 11 outputs in the last layer. We tested each model on the same dataset and with the same methodology as the SpikingJelly software model. 

The largest of these spiking CNNs as deployed comprises a total of 109,615 neurons, occupying approximately 86\% of a single-core system and thus approximately 0.07\% of the anticipated maximum system capacity.

Table \ref{tab:model_results} and Fig. \ref{fig:dvs_gesture_result} show the results. The match between the converted HiAER-Spike and postquantization SpikingJelly software test accuracies for the Spiking CNN variants ranged from perfect to a 1.74\% discrepancy in a model with two convolutional layers, each having ten output channels (included in Fig. \ref{fig:dvs_gesture_result} but not Table \ref{tab:model_results}). Adding more convolutional layers and more channels led to better test accuracies at the cost of higher energy consumption and latency per gesture instance.

Third, we trained a 3-layer CNN on 15-channel, 32 x 32 binarized CIFAR-10 images generated by bit-slicing. The architecture consists of three convolutional layers (3 x 3 kernels) with 16, 100, and 100 filters, respectively, followed by two linear layers (512 and 10 units). All convolutional layers use ReLU activation. The network was trained using cutout augmentation and PyTorch data augmentation techniques: RandomCrop, RandomHorizontalFlip, and AutoAugment. During training, an Adam optimizer with a weight decay of 0.05 and a cosine annealing scheduler that decayed from $1 \times 10^{-3}$ to $1 \times 10^{-7}$ over 200 epochs were used. We used a batch size of 64 and an early-stopping patience of 50 epochs. After training, the ANN was converted into an SNN using the SpikingJelly ann2snn converter, which replaces the ReLU activation in an ANN model with the IFNode of an SNN. We replaced the IFNode layers with the same custom spiking layers used in the DVS Gesture models and inserted a dropout regularization layer with p = 0.5 between the two linear layers. We fine-tuned the SNN using an ATan surrogate gradient function and the same data augmentation techniques and training hyperparameters as the ANN. The weights were then quantized to 16-bit integers and the network was converted to run on HiAER-Spike. The output neuron with the highest spike rate determined the predicted class. Table \ref{tab:model_results} shows the results.

Fourth, we implemented a deep Q-network (DQN) reinforcement learning agent to solve the Atari Pong task using a DVS-based input representation (Fig.~\ref{fig:pong}; \cite{deepmind2015}). In order to generate DVS events, our method compared each frame with the frame that was four frames prior to it. Standard RGB frames were converted to grayscale, downsampled, cropped, and converted into two event-based channels corresponding to binary ON and OFF pixel change events, at a spatial resolution of 84 × 84 pixels and with a change threshold of 10. The network architecture comprised three convolutional layers (2 input channels to 32 filters, 8 × 8 kernels, stride 4; 32 to 64 filters, 4 × 4 kernels, stride 2; and 64 to 64 filters, 3 × 3 kernels, stride 1), followed by flattening and two fully connected layers (3136 to 512 units, then 512 to 6 outputs corresponding to the six Pong actions).

Training followed standard DQN procedures with a learning rate of $1 \times 10^{-4}$, batch size of 64, replay buffer capacity of 400,000 transitions, and target network synchronization every 1000 updates. The trained PyTorch model was again converted to a spiking neural network (SNN) using the SpikingJelly ann2snn framework, replacing ReLU activations with integrate-and-fire (IFNode) layers. These IFNode layers were subsequently replaced with custom spiking layers of the same type that we used in our DVS Gesture and CIFAR-10 models (modeled after HiAER-Spike LIF neurons). The SNN was fine-tuned using ATan surrogate gradients on a dataset of the trained ANN playing Pong, consisting of 1 million frames of observations paired with corresponding ANN-selected actions. Rate coding over 20 simulation steps was employed to ensure stable Q-value estimation, with action selection determined by the maximum temporally accumulated membrane potential among the 6 outputs. The SNN underwent 16-bit quantization with dynamic alpha scaling to enable FPGA compatibility. Table \ref{tab:model_results} shows the results.

Finally, the performance of HiAER-Spike on MNIST as a basic benchmark and the DVS Gesture dataset as a more realistic benchmark was compared to other large-scale neuromorphic hardware in Table \ref{tab:mnistcomp} and Table \ref{tab:dvscomp}, respectively. Our results for HiAER-Spike shown here generally achieve lower accuracy than the other platforms showcased, but with superior energy and latency metrics. Our primary goal here was to test and demonstrate our HiAER-Spike FPGA hardware design, software interface, and conversion methods rather than to optimize models for accuracy. Optimization of model architectures, training methods (including data augmentation and regularization), quantization methods, and hyperparameters will lead to higher accuracy. In most instances, where reported, other platforms used a significantly greater number of time steps per inference, which tends to increase the accuracy of SNNs. Our MNIST results for HiAER-Spike are from neural networks consisting of ANN (binary) neurons rather than SNNs consisting of IF or LIF neurons. The DVS gesture results presented for HiAER-Spike use binarized and downsampled inputs, (2, 63, 63) and (2, 90, 90), rather than the full (2, 128, 128) inputs. Our choice to accumulate DVS data into 10 frames and then binarize the pixels may be improvable by choosing a more optimal method for accumulating events into frames.

Scaling was also analyzed within a single core. A linear regression was performed for the energy and latency metrics for the MLP, LeNet-5, and DVS Gesture spiking CNN networks across various neuron counts. Details on the regressions of the MLP and LeNet-5 data can be found in the Supplementary Information. The linear regressions for the DVS Gesture spiking CNN networks, for which we obtained scaling data across the largest range of neuron numbers, are $Energy (\mu J) = 0.0294x-30.293$ ($R^2 = 0.994$) and $Latency(\mu s) = 0.0658x-53.031$ ($R^2 = 0.995$), where $x$ is the number of neurons. While these high $R^2$ values suggest a strong fit, it is important to note that these observations are limited to a small sample size ($n=5$). Further investigation with a larger dataset would be necessary to confirm the observed linear trends. We are designing the hardware and software to simulate sparse networks and minimize the number of events transmitted over the higher-latency interconnects between cores, FPGAs, and servers (thereby keeping the majority of the event traffic on the faster on-chip routing connections). Consequently, we anticipate only modest deviations from these linear trends, due to the higher-latency connections, as the system scales to multiple cores, FPGAs, and servers. 

\section{Conclusions}

We presented a large-scale FPGA-based high-throughput neuromorphic SNN accelerator platform designed to serve as a shared resource for the neuromorphic computing research community specifically and STEM researchers more generally. We described the hardware stack as well as the software interface that allow users to configure and run a broad class of SNNs remotely on the system. Users are invited to submit jobs
over the NSG portal in the form of a simple Python script, and encouraged to provide feedback and request (or add) new features for the ongoing development of the system.

First experiments on HiAER-Spike show proof of concept that a single core of the system can run a relatively large spiking neural network for gesture recognition using DVS cameras with low latency and power consumption. Scaling analysis suggests a linear trend in increases in both latency and energy usage metrics, indicating the capability to scale performance to a much larger planned iteration of the system. 

Future work will extend the HiAER-Spike platform in several directions. On the one hand, we intend to make the system more versatile, introducing further properties of interest for computational neuroscience modeling, such as more sophisticated neuron models and learning rules. On the other, we strive to improve performance and efficiency for more machine-learning oriented applications such as on-line adaptive pattern recognition. 
We believe that this approach can bring together both computational neuroscience and artificial intelligence communities that have traditionally pursued disparate computational approaches.

\section*{Funding Declaration}

This work has been supported by National Science Foundation CNS-1823366 (CRI: CI-NEW: Trainable Reconfigurable Development Platform for Large-Scale Neuromorphic Cognitive Computing) and OAC-2346527 (CIRC: Grand: The Neuromorphic Commons (THOR)) and CCF-2208771 (Collaborative Research: FET: Medium: Energy-Efficient Persistent Learning-in-Memory with Quantum Tunneling Dynamic Synapses), Office of Naval Research N00014-20-1-2405 (Science of AI Brain Inspired Next Generation Deep Learning: Efficient and Persistent Online Learning with Spikes) and N00014-24-1-2127 (Science of AI: Massively Parallel Integrated CMOS-RRAM Arrays with Embedded Online Learning for AI on the Edge) and N00014-23-1-2162 (DURIP: High-Density Hybridized CMOS-RRAM Integration and Low-Noise Characterization of Large-Scale Neuromorphic Compute-in-Memory Arrays), and Western Digital Corporation.

\section*{Acknowledgments}

  We thank Sankar Basu, Mitchell Fream, Kameron Gano, Justin Kinney, Duygu Kuzum, Tim Liu, Martin Lueker-Boden, Amit Majumdar, Justin Mauger, Thomas McKenna, Nishant Mysore, Emre Neftci, Bruno Pedroni, Terrence Sejnowski, Dejan Vucini\'c, Riley Zeller-Townson, and organizers and participants of the Telluride Workshop on Neuromorphic Cognition Engineering for key and insightful contributions in the development and application of the HiAER-Spike CRI. We also thank Tomas Whitlock, Kevin Roth, and Alexandros Kapouranis at Alpha Data Inc. for extensive FPGA applications advice, Roger Miller at Samtec Corp. for FireFly help, Andrew Nelson at EXXACT Corp. for server help, Steven King at Arista for networking advice and Arista for donated switch hardware, Thomas Hutton and Christopher Cox at UCSD for switch configuration, Thomas Tate and all operations support staff at San Diego Supercomputer Center (SDSC) for installation help, Robert Buffington for IT support, Christopher Hughes of SDSC's High-Performance Computing group for networking support, and Andrew Ferbert, Ryan Nakashima, and William Homan of SDSC's Research Data Services (RDS) division for server support and admin.  

\section*{Author Contributions}

G.C., S.D., Q.W., L.G., and K.Y. contributed the concept for the HiAER-Spike community resource infrastructure for large-scale neuromorphic computing.  C.D., D.V., G.F., K.A., K.W., A.U., and M.Y. contributed to the software and user interface.  G.H. and O.O. developed and compiled the FPGA implementation. K.Y., L.G., Q.W., S.D., and G.C. contributed to the design and construction of the hardware at the San Diego Supercomputer Center and integration with the Neuroscience Gateway.  C.D., D.V., K.A., J.L., and L.G. contributed to the example applications and benchmarks.  All authors wrote and proofread the manuscript.14

\section*{Data Availability}
The MNIST dataset used in this study is publicly available from multiple sources, including the original source on Yann LeCun's website. The DVS Gesture dataset used in this study is publicly available and can be accessed from multiple sources, including Kaggle and various Python packages such as Tonic. The CIFAR-10 dataset is also available from multiple sources, including Kaggle and Alex Krizhevsky's CIFAR homepage on the University of Toronto Department of Computer Science website. Additional data for this study may be made available to qualified researchers on reasonable request from the corresponding authors.

\section*{Code Availability}
The hs\_api software package is available at \url{https://github.com/Integrated-Systems-Neuroengineering/hs_api/tree/main}. Additional underlying code for this study may be made available to qualified researchers on reasonable request from the corresponding authors.

\section*{Competing Interests}
All authors declare no financial or non-financial competing interests.

\FloatBarrier

\bibliography{refs}

@article{herculano2006cellular,
  title={Cellular scaling rules for rodent brains},
  author={Herculano-Houzel, Suzana and Mota, Bruno and Lent, Roberto},
  journal={Proceedings of the National Academy of Sciences},
  volume={103},
  number={32},
  pages={12138--12143},
  year={2006},
  publisher={National Acad Sciences}
}

@article{pehle2022brainscales,
  title={The BrainScaleS-2 accelerated neuromorphic system with hybrid plasticity},
  author={Pehle, Christian and Billaudelle, Sebastian and Cramer, Benjamin and Kaiser, Jakob and Schreiber, Korbinian and Stradmann, Yannik and Weis, Johannes and Leibfried, Aron and M{\"u}ller, Eric and Schemmel, Johannes},
  journal={Frontiers in Neuroscience},
  volume={16},
  pages={795876},
  year={2022},
  publisher={Frontiers}
}

@article{deepmind2015,
  title={Human-level control through deep reinforcement learning},
  author={Mnih, Volodymyr and Kavukcuoglu, Koray and Silver, David and Rusu, Andrei A and Veness, Joel and Bellemare, Marc G and Graves, Alex and Riedmiller, Martin and Fidjeland, Andreas K and Ostrovski, Georg and others},
  journal={nature},
  volume={518},
  number={7540},
  pages={529--533},
  year={2015},
  publisher={Nature Publishing Group}
}

@inproceedings{massa2020efficient,
  title={An efficient spiking neural network for recognizing gestures with a dvs camera on the loihi neuromorphic processor},
  author={Massa, Riccardo and Marchisio, Alberto and Martina, Maurizio and Shafique, Muhammad},
  booktitle={2020 International Joint Conference on Neural Networks (IJCNN)},
  pages={1--9},
  year={2020},
  organization={IEEE}
}

@inproceedings{arfa2025efficient,
  title={Efficient deployment of spiking neural networks on spinnaker2 for dvs gesture recognition using neuromorphic intermediate representation},
  author={Arfa, Sirine and Vogginger, Bernhard and Liu, Chen and Partzsch, Johannes and Sch{\"o}ne, Mark and Mayr, Christian},
  booktitle={2025 Neuro Inspired Computational Elements (NICE)},
  pages={1--8},
  year={2025},
  organization={IEEE}
}

@article{lenz2023ultra,
  title={Ultra-low-power Image Classification on Neuromorphic Hardware},
  author={Lenz, Gregor and Orchard, Garrick and Sheik, Sadique},
  journal={arXiv preprint arXiv:2309.16795},
  year={2023}
}

@article{esser2015backpropagation,
  title={Backpropagation for energy-efficient neuromorphic computing},
  author={Esser, Steve K and Appuswamy, Rathinakumar and Merolla, Paul and Arthur, John V and Modha, Dharmendra S},
  journal={Advances in neural information processing systems},
  volume={28},
  year={2015}
}

@inproceedings{stromatias2015scalable,
  title={Scalable energy-efficient, low-latency implementations of trained spiking deep belief networks on spinnaker},
  author={Stromatias, Evangelos and Neil, Daniel and Galluppi, Francesco and Pfeiffer, Michael and Liu, Shih-Chii and Furber, Steve},
  booktitle={2015 International Joint Conference on Neural Networks (IJCNN)},
  pages={1--8},
  year={2015},
  organization={IEEE}
}

@inproceedings{painkras2012spinnaker,
  title={Spinnaker: A multi-core system-on-chip for massively-parallel neural net simulation},
  author={Painkras, Eustace and Plana, Luis A and Garside, Jim and Temple, Steve and Davidson, Simon and Pepper, Jeffrey and Clark, David and Patterson, Cameron and Furber, Steve},
  booktitle={Proceedings of the IEEE 2012 Custom Integrated Circuits Conference},
  pages={1--4},
  year={2012},
  organization={IEEE}
}

@article{modha2023neural,
  title={Neural inference at the frontier of energy, space, and time},
  author={Modha, Dharmendra S and Akopyan, Filipp and Andreopoulos, Alexander and Appuswamy, Rathinakumar and Arthur, John V and Cassidy, Andrew S and Datta, Pallab and DeBole, Michael V and Esser, Steven K and Otero, Carlos Ortega and others},
  journal={Science},
  volume={382},
  number={6668},
  pages={329--335},
  year={2023},
  publisher={American Association for the Advancement of Science}
}

@article{davies2021taking,
  title={Taking neuromorphic computing to the next level with Loihi2},
  author={Davies, Mike and others},
  journal={Intel Labs’ Loihi},
  volume={2},
  pages={1--7},
  year={2021}
}

@inproceedings{sivagnanam_neuroscience_2020,
	location = {New York, {NY}, {USA}},
	title = {Neuroscience Gateway Enabling Large Scale Modeling and Data Processing in Neuroscience Research},
	isbn = {978-1-4503-6689-2},
	url = {https://dl.acm.org/doi/10.1145/3311790.3399625},
	doi = {10.1145/3311790.3399625},
	series = {{PEARC} '20},
	pages = {510--513},
	booktitle = {Practice and Experience in Advanced Research Computing},
	publisher = {Association for Computing Machinery},
	author = {Sivagnanam, Subhashini and Yoshimoto, Kenneth and Carnevale, Ted and Nadeau, Dave and Kandes, Martin and Petersen, Trevor and Truong, Dung and Martinez, Ramon and Delorme, Arnaud and Makeig, Scott and Majumdar, Amit},
	urldate = {2023-04-03},
	date = {2020-07-26},
     year="222",
}

@article{park2016hierarchical,
  title={Hierarchical address event routing for reconfigurable large-scale neuromorphic systems},
  author={Park, Jongkil and Yu, Theodore and Joshi, Siddharth and Maier, Christoph and Cauwenberghs, Gert},
  journal={IEEE transactions on neural networks and learning systems},
  volume={28},
  number={10},
  pages={2408--2422},
  year={2016},
  publisher={IEEE}
}

@article{mysore_hierarchical_2022,
	title = {Hierarchical Network Connectivity and Partitioning for Reconfigurable Large-Scale Neuromorphic Systems},
	volume = {15},
	issn = {1662-453X},
	url = {https://www.frontiersin.org/articles/10.3389/fnins.2021.797654},
	journaltitle = {Frontiers in Neuroscience},
	author = {Mysore, Nishant and Hota, Gopabandhu and Deiss, Stephen R. and Pedroni, Bruno U. and Cauwenberghs, Gert},
	urldate = {2023-04-03},
	date = {2022},
     year="222",
}

@article{correll2020fully,
  title={A fully integrated reprogrammable CMOS-RRAM compute-in-memory coprocessor for neuromorphic applications},
  author={Correll, Justin M and Bothra, Vishishtha and Cai, Fuxi and Lim, Yong and Lee, Seung Hwan and Lee, Seungjong and Lu, Wei D and Zhang, Zhengya and Flynn, Michael P},
  journal={IEEE Journal on Exploratory Solid-State Computational Devices and Circuits},
  volume={6},
  number={1},
  pages={36--44},
  year={2020},
  publisher={IEEE}
}

@inproceedings{kulkarni2019neuromorphic,
  title={Neuromorphic hardware accelerator for SNN inference based on STT-RAM crossbar arrays},
  author={Kulkarni, Shruti R and Kadetotad, Deepak Vinayak and Yin, Shihui and Seo, Jae-Sun and Rajendran, Bipin},
  booktitle={2019 26th IEEE International Conference on Electronics, Circuits and Systems (ICECS)},
  pages={438--441},
  year={2019},
  organization={IEEE}
}

@article{pedroni2019memory,
  title={Memory-efficient synaptic connectivity for spike-timing-dependent plasticity},
  author={Pedroni, Bruno U and Joshi, Siddharth and Deiss, Stephen R and Sheik, Sadique and Detorakis, Georgios and Paul, Somnath and Augustine, Charles and Neftci, Emre O and Cauwenberghs, Gert},
  journal={Frontiers in neuroscience},
  volume={13},
  pages={357},
  year={2019},
  publisher={Frontiers Media SA}
}

@INPROCEEDINGS{iscas_hiearahb,
  author={Hota, Gopabandhu and Mysore, Nishant and Deiss, Stephen and Pedroni, Bruno and Cauwenberghs, Gert},
  booktitle={2022 IEEE International Symposium on Circuits and Systems (ISCAS)}, 
  title={Hierarchical Multicast Network-On-Chip for Scalable Reconfigurable Neuromorphic Systems}, 
  year={2022},
  volume={},
  number={},
  pages={481-485},
  doi={10.1109/ISCAS48785.2022.9937961}}

@article{deng2020rethinking,
  title={Rethinking the performance comparison between SNNS and ANNS},
  author={Deng, Lei and Wu, Yujie and Hu, Xing and Liang, Ling and Ding, Yufei and Li, Guoqi and Zhao, Guangshe and Li, Peng and Xie, Yuan},
  journal={Neural networks},
  volume={121},
  pages={294--307},
  year={2020},
  publisher={Elsevier}
}

@inproceedings{amir2017low,
  title={A low power, fully event-based gesture recognition system},
  author={Amir, Arnon and Taba, Brian and Berg, David and Melano, Timothy and McKinstry, Jeffrey and Di Nolfo, Carmelo and Nayak, Tapan and Andreopoulos, Alexander and Garreau, Guillaume and Mendoza, Marcela and others},
  booktitle={Proceedings of the IEEE Conference on Computer Vision and Pattern Recognition},
  pages={7243--7252},
  year={2017}
}

\appendix

\section{Supplementary Information}

\subsection{API Definition}
\begin{figure}
\begin{center}
\textbf{Example Network}\par\medskip
  \includegraphics[scale=0.12]{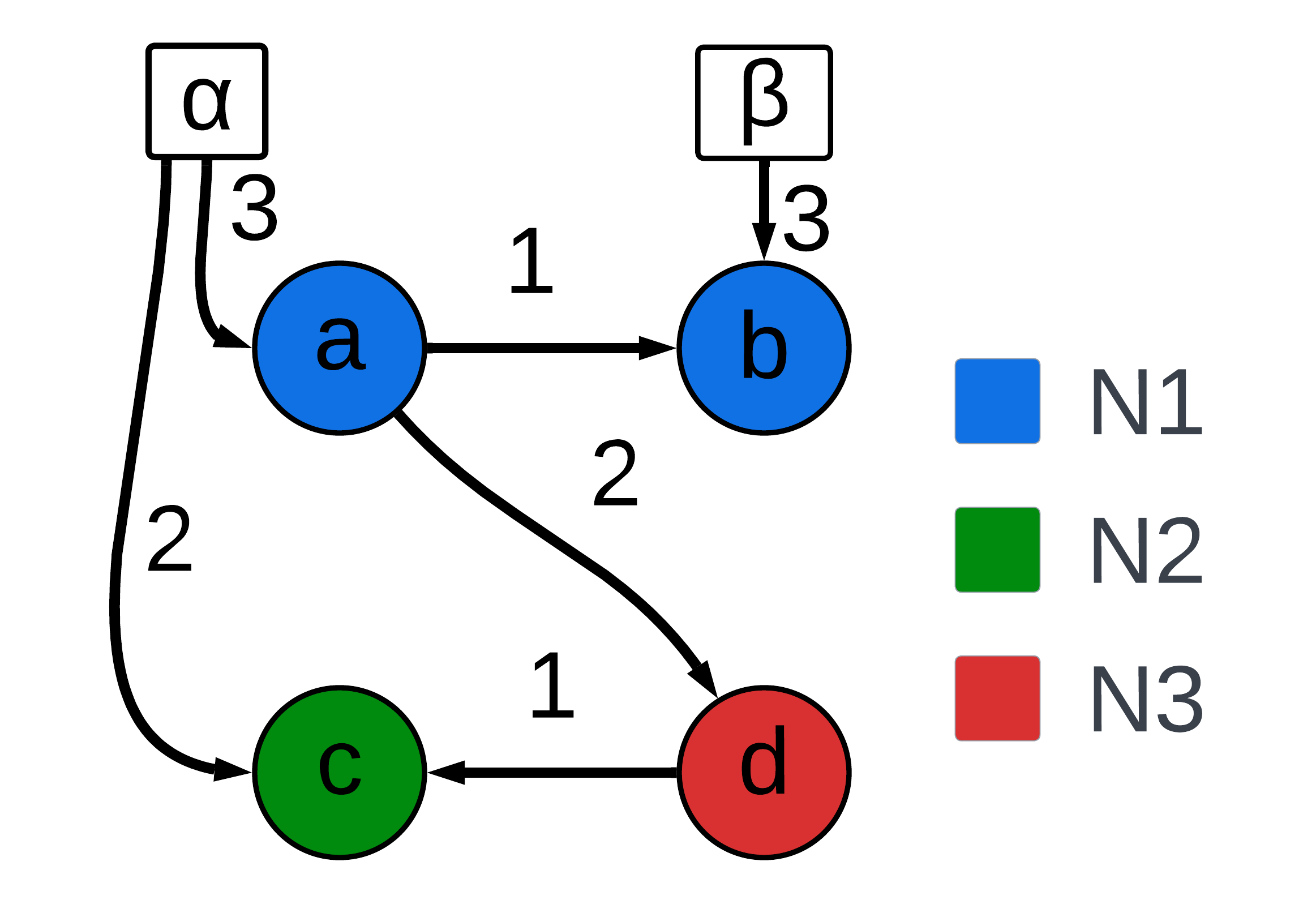}
    \begin{lstlisting}[language=Python, frame=lines, framesep=2mm, basicstyle=\footnotesize\ttfamily]
from hs_api.api import CRI_network
from hs_api.neuron_models import LIF_neuron, ANN_neuron

N1 = LIF_neuron(theta = 3, nu = -17, Lambda = (2**6)-1)
N2 = LIF_neuron(theta = 4, nu = -17, Lambda = 2)
N3 = ANN_neuron(theta = 5, nu = 2)

axons = {'alpha': [('a', 3),('c', 2)],
             'beta': [('b', 3)]}

neurons = {'a': ([('b', 1), ('d', 2)], N1),
                   'b': ([], N1),
                   'c': ([], N2),
                   'd': ([('c', 1)], N3)}

outputs = ['a', 'b']

network = CRI_network(axons=axons,
    neurons=neurons,
    outputs=outputs)
inputs = ['alpha','beta']
currSpikes = network.step(inputs)
  \end{lstlisting}
  \caption{Example network and generating code.}

  \label{fig:examp-network}
\end{center}
\end{figure}

This section demonstrates how to create the network shown in Fig. \ref{fig:examp-network}, which consists of four neurons, $a$ through $d$, and two axons, $\alpha$ and $\beta$. In the example network, neurons $a$ and $b$ are leaky-integrate-and-fire neurons with no membrane potential noise and almost no leak (achieved by setting a large $\lambda$ value) with a threshold of 3, neuron $c$ is a leaky-integrate-and-fire neuron with no membrane potential noise ($\mu$) and a leak ($\lambda$) parameter of 2 and a threshold ($\theta$) of 4, and neuron $d$ is an ANN neuron with a threshold of 5 with membrane potential noise. A simple network was intentionally selected for demonstration purposes.

First, the user must define a set of  neuron models:
\begin{lstlisting}[language=Python]
N1 = LIF_neuron(theta = 3,
    nu = -17, Lambda = (2**6)-1)
N2 = LIF_neuron(theta = 4,
    nu = -17, Lambda = 2)
N3 = ANN_neuron(theta = 5,
    nu = 2)
\end{lstlisting}

Each neuron is an instance of the \lstinline[language=Python]{LIF_neuron} or \lstinline[language=Python]{ANN_neuron} class. Users can create many neuron models with different parameters. Neuron models can be assigned to any neuron in the network, but each neuron must be assigned only one neuron model.

Next, the user must supply an \lstinline[language=Python]{axons} dictionary:
\begin{lstlisting}[language=Python]
axons = {'alpha': [('a', 3),('c', 2)],
             'beta': [('b', 3)]}
\end{lstlisting}

Axons represent user-controllable input from the external world into the network through which a user may send a spike to interact with multiple postsynaptic neurons. The keys in this dictionary are unique Python objects, usually strings, representing each axon coming into the network. The values associated with these keys are lists of tuples specifying the respective synaptic connections to one or more neurons in the network. Each tuple contains two elements, the postsynaptic neuron's unique key and an integer representing the weight of the connection. In this case, we create the \lstinline[language=Python]{"alpha"} axon with synapses to the \lstinline[language=Python]{"a"} and \lstinline[language=Python]{"c"} neurons with weights of \lstinline[language=Python]{3} and \lstinline[language=Python]{2}, respectively, and the \lstinline[language=Python]{"beta"} neuron with a synapse to the \lstinline[language=Python]{"b"} neuron with a weight of \lstinline[language=Python]{3}.

The user must also supply a \lstinline[language=Python]{neurons} dictionary:
\begin{lstlisting}[language=Python]
neurons = {'a': ([('b', 1), ('d', 2)], N1),
                   'b': ([], N1),
                   'c': ([], N2),
                   'd': ([('c', 1)], N3)}
\end{lstlisting}

The keys in this dictionary are unique objects that represent each neuron in the network, usually a number or string. The corresponding values are tuples where the first element is a list of all of the neurons' outgoing synapses and the second element is a neuron model used to specify membrane updates for the neurons. Each outgoing synapse is represented by a tuple comprising the postsynaptic neuron's unique key and the integer weight of the synapse. In this case, we instantiate the \lstinline[language=Python]{"a"} neuron with synapses to the \lstinline[language=Python]{"b"} and \lstinline[language=Python]{"d"} neurons with weights of \lstinline[language=Python]{1} and \lstinline[language=Python]{2} respectively, neurons \lstinline[language=Python]{"b"} and \lstinline[language=Python]{"c"} without outgoing synaptic connections, and the \lstinline[language=Python]{"d"} neuron with synapses to the neuron \lstinline[language=Python]{"c"} with a weight of \lstinline[language=Python]{1}.

Finally, the user must supply an outputs list:
\begin{lstlisting}[language=Python]
outputs = ['a', 'b']
\end{lstlisting}

This is a list of all the neurons' unique keys whose spiking activity the user wishes to monitor. In this case we designate neurons \lstinline[language=Python]{"a"} and \lstinline[language=Python]{"b"} as output neurons.

Once the necessary data structures are defined, they may be passed to the constructor to create an instance of the 
\lstinline[language=Python]{CRI_network} class that exposes functions through which the user can interact with the network:

\begin{lstlisting}[language=Python]
network = CRI_network(
    axons=axons,
    neurons=neurons,
    outputs=outputs)
inputs = ['alpha','beta']
currSpikes = network.step(inputs)
\end{lstlisting}

The \lstinline[language=Python]{step} method can be used to run one timestep of the network. The user supplies an inputs list consisting of the unique keys of each axon the user wishes to activate during the timestep. In this case, both axons \lstinline[language=Python]{"alpha"} and \lstinline[language=Python]{"beta"} are activated. The function returns a list of all output neurons that spiked during the timestep. If the optional \lstinline[language=Python]{membranePotential} flag is set, the function also returns the membrane potential for every neuron in the network.

The methods \lstinline[language=Python]{read_synapse} and \lstinline[language=Python]{write_synapse} can be used to change synapse weights after calling the 
\lstinline[language=Python]{CRI_network} constructor. The \lstinline[language=Python]{read_synapse} method takes the unique key for the presynaptic axon or neuron and the postsynaptic neuron and returns the corresponding synapse weight. The \lstinline[language=Python]{write_synapse} method, accordingly, takes two keys and a new weight to set for the synapse. Here, we just increment the weight of the synapse from neuron \lstinline[language=Python]{"a"} to \lstinline[language=Python]{"b"} by one:
\begin{lstlisting}[language=Python]
currWeight = network.read_synapse('a', 'b')
network.write_synapse('a', 'b',
    currWeight+1)
\end{lstlisting}

The \lstinline[language=Python]{read_membrane} method takes a list of the unique keys for a set of neurons and returns a list of the corresponding membrane potentials for those neurons.
\begin{lstlisting}[language=Python]
MPs = network.read_membrane('a', 'b')
\end{lstlisting}

\subsection{Converting PyTorch models}\label{converting-pytorch}

Converting PyTorch models to run on the HiAER-Spike system involves representing inputs as axon activations. Images are binarized into one or several channels with an axon representing each pixel. Each axon is inserted into the axon dictionary. To pass an image into the network, we append the keys of axons corresponding to each active pixel to the input list. Passing the input list into the network at a time step activates the unique set of axons for each image.

The axons are connected to the model's first layer, which, for many of the vision models in this paper, is a convolutional layer. The pixels of the output feature maps are represented using neurons. To determine the postsynaptic neurons of each axon, we implemented a mapping technique utilizing a PyTorch tensor with the same dimensions as the input image. The tensor is filled in row-major order with consecutive integers starting from 0, labeling each pixel location with the index of its corresponding axon. A window slides over the tensor, mimicking the movements of the layer's kernel, to identify the axons connected to each feature map neuron. For each axon in the window's current position, a unique tuple containing the postsynaptic neuron and synaptic weight is appended to the axon's list of outgoing synapses. Every postsynaptic neuron is inserted into the neuron dictionary with a key containing the feature map it belongs to and its index within that map. We repeat this process for each learnable kernel in the convolutional layer.

The neurons representing the convolutional layer are connected to the subsequent layer. If the next layer is another convolutional layer, we use the same mapping technique to identify the postsynaptic neurons representing the second convolutional layer. The PyTorch tensor has the same dimensions as the output feature maps from the first convolutional layer. For each neuron in the window's current position, a unique tuple containing the postsynaptic neuron and synaptic weight is appended to the neuron's list of outgoing synapses. Every postsynaptic neuron is inserted into the neuron dictionary. If the next layer is fully connected, each neuron in the convolutional layer has outgoing synapses to every neuron in the fully connected layer. We iterate through every neuron in the convolutional layer, appending a tuple containing its postsynaptic neuron and synaptic weight to its list of outgoing synapses. Every neuron in the fully connected layer is inserted into the neuron dictionary with a key containing its index within that layer.

There are several ways to implement biases in the HiAER-spike network. The first method is to simply subtract the bias from the threshold of the corresponding neuron. The second method involves connecting an axon to the neuron and setting its synaptic weight equal to the corresponding bias term. The neuron's threshold is unchanged. Alternatively, the axons can be replaced with ANN neurons with thresholds equal to -1. These neurons are always active due to their negative thresholds.

The neurons in the output layer determine the model's prediction. We append the keys of the output neurons to the output list. The network will report if any of the output neurons spike after each timestep. We can also pass the output list into the read\_membrane() function of the network to output their membrane potentials. These methods allow us to determine the model's prediction based on the spiking or membrane potential of the output neurons.

The conversion techniques can be repeated or modified to convert complex PyTorch models with varying topologies into HiAER-Spike networks.  The software interface allows end users to convert their models into organized data structures with transparent fan-in and fan-out lists. As a result, the converted model is easy to inspect, debug, and modify. Users can extract valuable information during inference, such as spiking events and membrane potentials, to determine model accuracy and behavior. The software provides a practical and reliable conversion pipeline for running PyTorch models on the HiAER-spike system.


\subsection{Managing HBM}\label{managing-hbm}

\begin{figure}
\begin{center}
 \textbf{Axon HBM Mapping}\par\medskip
  \includegraphics[scale=0.7]{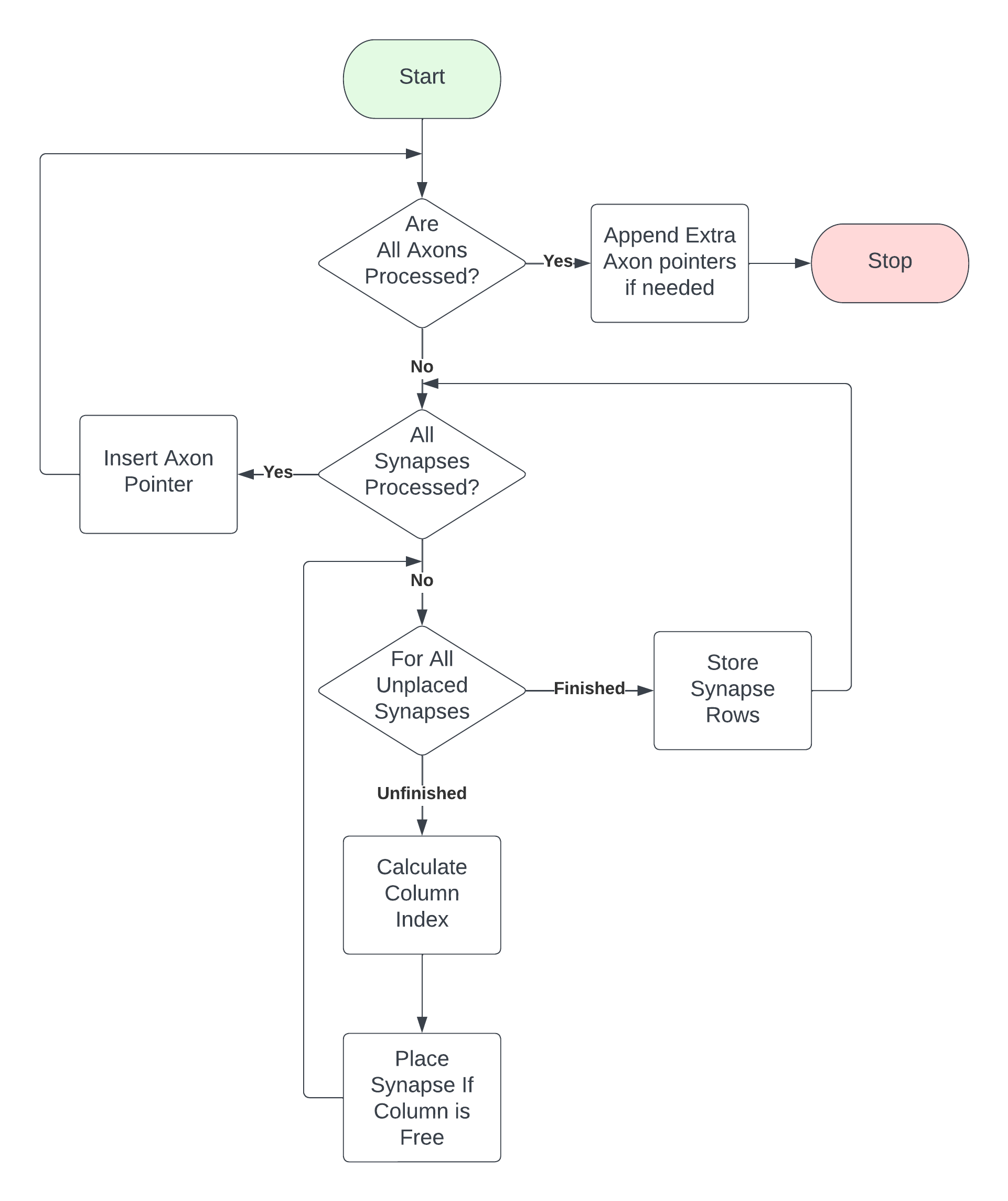}
  \caption{Simplified flowchart describing the process of mapping a network into the HBM synaptic routing table.}
  \label{fig:axon-flow}
\end{center}
\end{figure}

When running on the HiAER-Spike cluster, the software package must not only send commands to the FPGAs over PCIe to edit synapses, retrieve spike data, and orchestrate execution, but also initialize the network topology in HBM prior to execution. When a user creates a network, the software first partitions the network across all available cores/FPGAs/servers (if any) following the scheme described in [10]. After partitioning, the software maps the partitioned network into HBM for each core. Memory in HBM is divided into four sections, a section to hold neuron model definitions, a section to hold axon pointers, a section to hold neuron pointers, and a section to hold synapses.

Neuron models are programmed into HBM and point to specific sections of HBM that hold the neuron pointers for neurons assigned to the specific neuron model.
Axons are programmed into HBM according to the process described in (Fig.~\ref{fig:axon-flow}), which iterates through all axons and iterates through all synapses of each axon and assigns each a space in HBM.
All the synapses outgoing from an axon must be placed in a contiguous space in HBM. However, synapse definitions in HBM must be aligned with the address of their postsynaptic neuron. So each synapse for an axon is placed into memory following the needed alignment; and once all the synapse definitions for a given axon are written into HBM, an axon pointer is inserted into the axon pointer region of memory that points to the addresses in which the synapse definitions for that axon are stored.

Neurons are programmed into HBM in a similar manner, with three extra steps: First, neurons are grouped by their corresponding neuron models and assigned to a section of HBM reserved for neuron pointers of neurons with each specific neuron model. Second, in order to designate a neuron as an output neuron, a special flag must be set in the synapse definitions for that neuron. If necessary, the region in memory pointed to by a neuron can be expanded to accommodate this flag, by adding dummy synapses. Third, if a neuron has no outgoing synapses, a set of 16 zero-weight synapses are inserted into HBM so that every neuron has a space in the section of HBM that holds synapses.

\begin{figure}
\begin{center}
\textbf{Simulator Code}\par\medskip
\end{center}
\begin{lstlisting}[language=Python, frame=lines, framesep=2mm, basicstyle=\footnotesize\ttfamily]
# generate a random perturbation for the membrane
# potential of each neuron in the network
perturbBits = 17
perturbation = np.random.randint(
                    -1 * 2 ** (perturbBits - 1),
                    2 ** (perturbBits - 1),
                    size=numNeurons
                    )

# balancing the positive and negative distribution by setting LSB to 1
perturbation(perturbation | 1)
# shift the noise to be applied to each neuron according to the nu value
# assigned by that neuron's model. Left shift for positive values,
# right shift for negative
perturbation = leftshiftArr(perturbation, self.nuByModel,
                            np.greater(self.nuByModel, 0))
perturbation = rightshiftArr(perturbation, np.absolute(self.nuByModel),
                            np.less(self.nuByModel, 0))
# add the noise to the membrane potential
self.membranePotentials(self.membranePotentials + perturbation)
# check membrane potentials and determine spiked neurons
spiked_inds = np.nonzero(membranePotentials > threshold)
membranePotentials[spiked_inds] = 0
firedNeurons = np.transpose(spiked_inds).flatten().tolist()
# update LIF neurons
if lifNeurons.size > 0:
    self.membranePotentials[lifNeurons] = self.membranePotentials[lifNeurons]
    - (self.membranePotentials[lifNeurons] // np.power(2, Lambdas[lifNeurons]))
# update ANN neurons
if memLessNeurons.size > 0:
    self.membranePotentials[memLessNeurons] = 0
firedAxons = np.zeros(numAxons)
firedAxons[inputs] = 1
firedNeurons = np.zeros(numNeurons)
firedNeurons[spiked_inds] = 1
membraneUpdatesAxon = axonWeights @ firedAxons
membraneUpdates = neuronWeights @ firedNeurons
membranePotentials = membranePotentials + membraneUpdates
    + membraneUpdatesAxon
stepNum = stepNum+1
outputSpikes = [ i for i in self.firedNeurons if i in outputs]
\end{lstlisting}
\caption{Python code that emulates the FPGA hardware. The simulator can be used for developing code offline and can be easily substituted for the real data streams to and from the FPGA hardware over the PCIe interface. We use this emulation as a further benchmarking tool to compare the throughput of the FPGA implementation to a pure software implementation running on the CPU.}

  \label{fig:sim-code}
\end{figure}

\begin{figure}
\begin{center}
\textbf{HiAER-Spike Hardware}\par\medskip
  \includegraphics[width=\columnwidth]
  {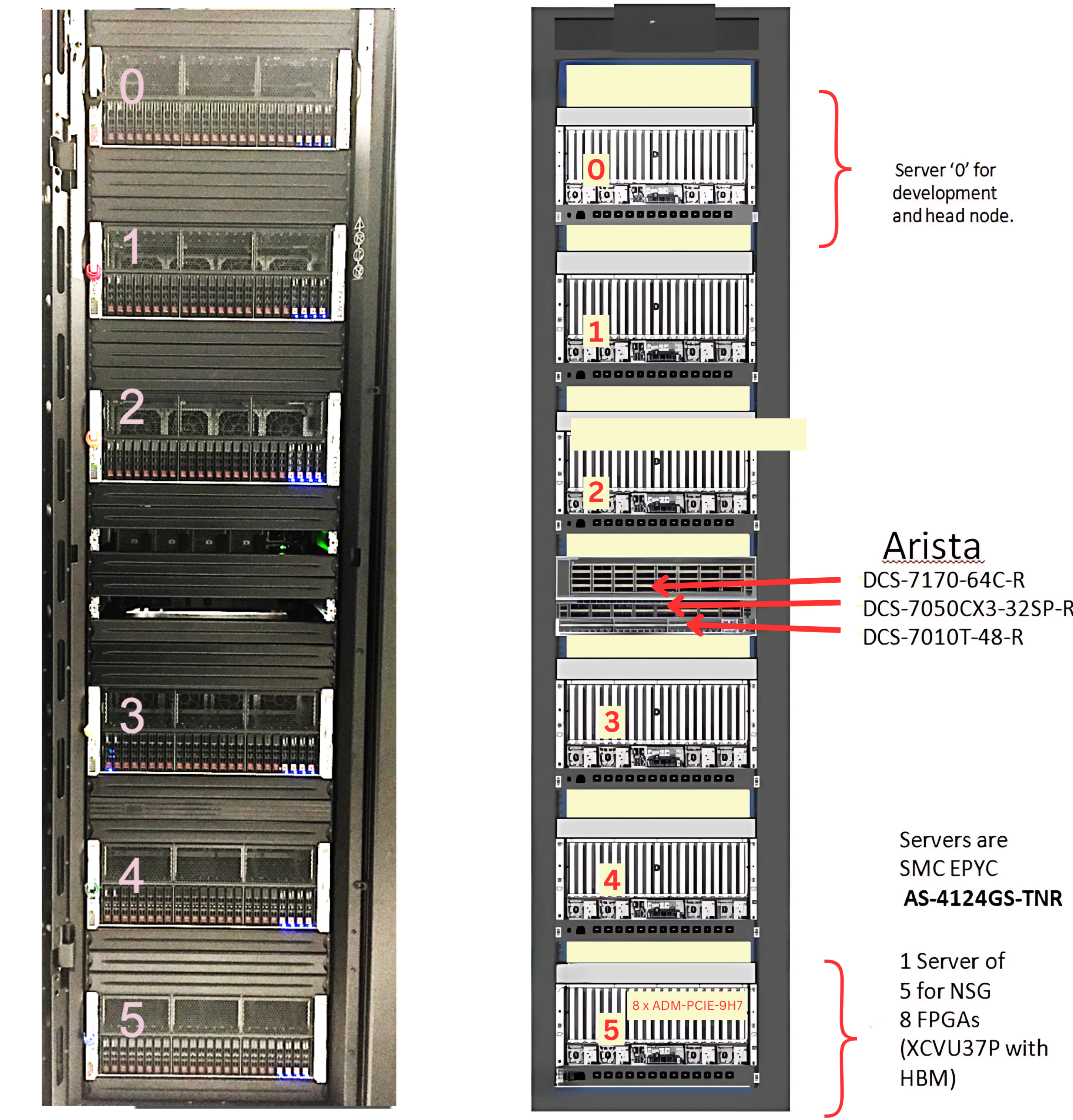}
  \caption{The HiAER-Spike system as implemented at the San Diego Supercomputer Center, consisting of 6 servers, one coordinating the system and the other five each containing 8 FPGAs. FPGAs within a server are connected together using FireFly cables, and FPGAs will be connected between servers using a custom Ethernet protocol. Three Arista switches service the system; the Arista 7170 will implement the custom Ethernet routing protocol.}
  \label{fig:rack}
\end{center}
\end{figure}

\begin{figure}
\begin{center}
\textbf{HiAER-Spike Scaling}\par\medskip
  \includegraphics[scale=0.4] {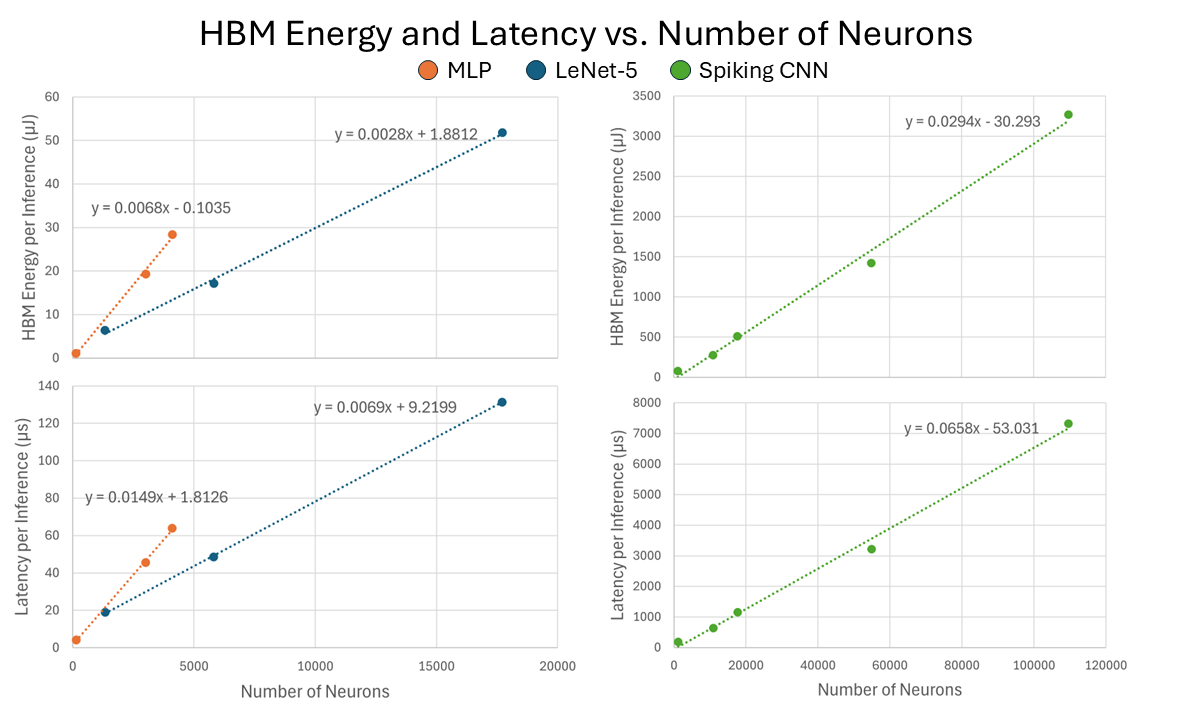}

  \caption{The relationship between HBM energy/latency per inference and the number of neurons for three different model types run on a single FPGA core: MLP, LeNet-5, and DVS Gesture spiking CNN. HBM energy/latency per inference scales linearly with the number of neurons in a model. The linear fits calculate the approximate increase in HBM energy/latency per inference for each additional neuron. The HBM energy and latency cost per neuron depends on the specific model architecture and type. The MLP models cost about 2.4 times more HBM energy and 2.2 times more latency per neuron than the LeNet-5, likely attributable to the larger number of connections per neuron (for most neurons) in MLP models. The DVS Gesture spiking CNN models cost about 10.5 times more HBM energy and 9.5 times more latency per neuron than the LeNet-5, likely attributable to the use of 10 frames per inference (requiring 10 timesteps) for the spiking CNN models.}

  \label{fig:scaling}
\end{center}
\end{figure}


\end{document}